\documentclass[letterpaper]{IEEEtran}

\usepackage{color}

\usepackage{subfigure}

\usepackage{graphicx}
\usepackage[cmex10]{amsmath}
\usepackage{amssymb, amsfonts, amsthm}
\usepackage{multicol}
\usepackage{multicol}
\usepackage{multirow}

\usepackage[bookmarks=false]{hyperref}
\usepackage{times}
\usepackage{courier}
\usepackage{helvet}
\usepackage{arial}

\newtheorem{theorem}{Theorem}
\newtheorem{lem}{Lemma}
\newtheorem{prop}{Proposition}
\newtheorem{corollary}{Corollary}

\IEEEoverridecommandlockouts

\begin{document}

\title{Entry and Spectrum Sharing Scheme Selection in Femtocell Markets$^{{\footnotesize\dagger}}$
\thanks{$^{{\footnotesize\dagger}}$ This work is supported in part by National Science Foundation under
Grant No. 0830556. S. Ren and M. van der Schaar are  with Electrical
Engineering Department, University of California, Los Angeles
(UCLA). E-mail: \{rsl,mihaela\}@ee.ucla.edu. J. Park was with
Electrical Engineering Department, UCLA, and is now with School of
Economics, Yonsei University. E-mail:  jaeok.park@yonsei.ac.kr. }}

\author{Shaolei~Ren,~\IEEEmembership{Student Member,~IEEE,} Jaeok Park,
and  Mihaela van der
Schaar,~\IEEEmembership{Fellow,~IEEE}}

\markboth{}%
{Ren, Park, \lowercase{and} van der Schaar: Entry and Spectrum Sharing Scheme Selection in Femtocell Communications Markets}

\maketitle

\begin{abstract}

Focusing on a femtocell communications market, we study  the entrant
network service provider's (NSP's) long-term decision: whether to
enter the market and which spectrum sharing technology to select to
maximize its profit. This long-term decision is closely related to
the entrant's pricing strategy and the users' aggregate demand,
which we model as medium-term and short-term decisions,
respectively. We consider two markets, one with no incumbent and the
other with one incumbent. For both markets, we show  the existence
and uniqueness of an equilibrium point in the user subscription
dynamics, and provide a sufficient condition for the convergence of
the dynamics. For the market with no incumbent, we derive upper and
lower bounds on the optimal price and market share that maximize the
entrant's revenue, based on which the entrant selects an available
technology to maximize its long-term profit. For the market with one
incumbent, we model competition between the two NSPs as a
non-cooperative game, in which the incumbent and the entrant choose
their market shares independently, and provide a sufficient
condition that guarantees the existence of at least one pure Nash
equilibrium. Finally, we formalize the problem of entry and spectrum
sharing scheme selection for the entrant and provide numerical
results to complement our analysis.

\end{abstract}


\section{Introduction}

Enhancing indoor wireless connectivity is a major challenge that
hinders the proliferation of future-generation wireless networks
operating at high frequencies, as signals at these frequencies
suffer from severe fading and attenuation. Recently, femtocells
(i.e., home base stations) have been proposed as an enabling
solution to improve the indoor wireless communications service in 4G
data networks \cite{HobbyClaussen}. Due to the wireless nature of
femtocells, spectrum management for the coexistence of femtocells
and macrocells is essential to realize the full potential of
femtocells, which will be a key factor in the successful adoption of
femtocells in the future communications market. Broadly speaking,
there
 are three  spectrum sharing schemes (or technologies) for the coexistence of
 femtocell and macrocell base stations \cite{HobbyClaussen}: (1) ``split'':
 part of the spectrum owned by the NSPs
 is dedicated to femtocells; (2) ``common'': the macrocell and the femtocells operate
on the same spectrum and hence interfere with each other; (3)
``partially shared'': the femtocells operate only on a fraction of
the spectrum used by the macrocells.
While these three spectrum sharing schemes have their respective
advantages (e.g., spectrum efficiency, interference level), there is an ongoing debate
over which scheme shall be adopted \cite{ShettyParekhWalrand}.

\begin{figure}[htp]
\begin{center}
\includegraphics[width=8.7 cm]{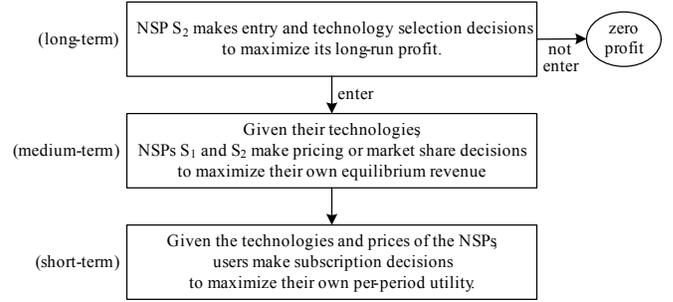}
\caption{\normalfont Three-stage decision-making problems of entrant
and incumbent NSPs.} \label{timing}
\end{center}
\end{figure}

Because of the potential of significantly improving indoor
communications services, femtocells are being adopted by more and
more NSPs and meanwhile, create new business opportunities for
start-ups which can enter the communications market by providing
femtocell services. Thus, it is important to investigate whether or
not it is profitable for an entrant NSP to enter a market with
femtocell services and with which technologies (e.g., which spectrum
sharing schemes). From an economics perspective, we study in this
paper the problem of entry and spectrum sharing scheme\footnote{We
implicitly assume that the entrant has decided in advance how much
bandwidth to acquire if it chooses to enter the market. This
assumption can be relaxed without affecting our analysis. The
decision of spectrum allocation (i.e., how much bandwidth allocated
to femtocells and macrocells) is not explicitly considered in the
paper. However, spectrum allocation decision can be captured if we
treat different spectrum allocations (but possibly with the same
spectrum sharing scheme) as different technologies in the set
$\{\mathcal{T}_0,\mathcal{T}_1,\mathcal{T}_2,\cdots
\mathcal{T}_L\}$. Please see Section III for more details.}
selection faced by a profit-maximizing entrant network service
provider (NSP) in a femtocells communications market. In particular,
our study shall quantitatively characterize which spectrum sharing
scheme shall be adopted by the entrant to maximize its profit and
under which conditions. Two markets, one with no incumbent and the
other with one incumbent, shall be investigated in this paper.
Throughout the paper, we  use ``spectrum sharing scheme'' and
``technology'' interchangeably wherever applicable. The structure of
our analysis is shown in Fig. \ref{timing}. Specifically, we
consider a three-stage decision making process: in the long term,
the entrant NSP, denoted by $\mathcal{S}_2$, makes entry and
technology selection decisions to maximize its long-run profit; in
the medium term, the incumbent NSP, denoted by $\mathcal{S}_1$, and
NSP $\mathcal{S}_2$ make pricing (or market share) decisions to
maximize their own revenue; and in the short term, users make
subscription decisions to maximize their their own per-period
utility. This three-stage hierarchical order of decision making can
be explained as follows. Although dynamic spectrum management for
femtocells has been proposed as a research thrust (e.g.,
\cite{LeeYongNguyenLee}), deployment of a spectrum sharing scheme
incurs a large cost, as it requires the network infrastructure and
femtocell terminals to be manufactured in compliance with the chosen
spectrum sharing scheme \cite{ShettyParekhWalrand}. It also requires
the support of protocol stacks, which is costly to develop. For
instance, if ``split'' is chosen, then the femtocells terminals
should be designed and manufactured such that they are only able to
operate on certain bandwidths dedicated to femtocells. Therefore,
the spectrum sharing scheme is difficult to change once deployed and
hence, it is a long-term strategy for an NSP
\cite{ShettyParekhWalrand}.\footnote{Note that once the long-term
spectrum sharing scheme is determined, dynamic spectrum management
in femtocells is still possible (e.g., dynamic frequency hopping
among femtocell users depending on certain criteria such as their
instantaneous channel conditions).}
 In contrast, an NSP can adjust its price over the lifespan of its
network infrastructure, although the price cannot be updated as
frequently as the users change their subscription decisions.
Overall, we can assume that the users may change their subscription
decisions frequently (e.g., a few days or weeks as a period), an
NSP's price is changed less frequently (e.g., several months or
years as a period), while an NSP's technology is a long-term
decision (e.g., several years as a period). In order to evaluate and
compare the long-term profitability of networks with different
technologies, the entrant NSP needs to predict its maximum profit
for each available technology. To maximize revenue given the
technology and the associated cost, the NSP needs to know the users'
aggregate demand and their willingness to pay for the service, and
then choose their optimal prices.
 Hence, we study first users' dynamic decisions as to
whether or not they subscribe to the entrant for communications
services (i.e., the short-term problem), then revenue-maximizing
pricing strategies (i.e., the medium-term problem), and finally
entry and technology selection for the potential entrant (i.e., the
long-term problem). A similar hierarchical analysis was considered
in \cite{ShettySchwartzWalrand} in the contexts of Internet markets.
Note that our study of user subscription dynamics provides a deeper
understanding of the users' subscription decisions than directly
assuming a certain form of demand function (e.g.,
\cite{ShettySchwartzWalrand}), since our study characterizes both
the dynamics and equilibrium point in the process of users'
subscription decisions.

When more users share the same network infrastructure, congestion
effects are typically observed in communications networks and
especially in wireless networks where interferences add to the
difficulty in spectrum management
\cite{AcemogluOzdaglar}\cite{ZemlianovVeciana}. In economics terms,
congestion effects can be classified as a type of negative network
externalities \cite{Tirole}. To capture congestion effects, we
assume that the entrant provides each user with a QoS
 which is modeled as a non-increasing function in terms of the
number of subscribers. In the first part of this paper, we focus on
a market with no incumbent.
 By jointly considering the provided QoS and the charged
price, users can dynamically decide whether or not to subscribe to
the entrant. Under the assumption that users make their subscription
decisions based on the most recent QoS and the current price, we
show that, for any QoS function and price, there exists a unique
equilibrium point of the user subscription dynamics at which the
number of subscribers does not change. Given a spectrum sharing
scheme, if the QoS
degrades too fast when more users subscribe to the entrant, the user
subscription dynamics
 may not  converge to the equilibrium point. Hence, we find a sufficient condition
for the QoS function to
 ensure the global convergence of the user
subscription dynamics. We also derive upper and lower bounds on the
optimal price and market share that maximize the entrant's revenue.
With linearly-degrading QoS functions (which we show can approximate
very well the actual QoS functions), we obtain the optimal price in
a closed form. Then, the entrant can select a technology out of its
available options such that its long-term profit is maximized. Next,
we turn to the analysis of a market with one incumbent in the second
part of our paper. For the convenience of analysis, we assume that
the incumbent
 has sufficient resources to provide each subscriber
a guaranteed QoS (or to be more precise, a QoS that degrades
sufficiently slowly such that it can be approximated as a constant
without losing much accuracy).\footnote{The assumption of a constant
QoS is relaxed in the subsequent work
\cite{RenVanderschaarTechnicalReport}, where we focus the capacity
investment decision and consider heterogeneity in the users' data
demands as well.} Given the provided QoS and the charged prices,
users dynamically select the NSP that yields a higher (positive)
utility. We first show that, for any prices, the considered user
subscription dynamics always admits a unique equilibrium point, at
which no user wishes to change its subscription decision. We next
obtain a sufficient condition for the QoS functions to guarantee the
convergence of
the user subscription dynamics. 
Then, we analyze  competition between the two NSPs. Specifically,
modeled as a strategic player in a non-cooperative game, each NSP
aims to maximize its own revenue by selecting its own market share
while regarding the market share of its competitor as fixed. This is
in sharp contrast with the existing related literature which
typically studies  price competition among NSPs. For the formulated
market share competition game, we  derive a sufficient condition on
the QoS function that guarantees the existence of at least one pure
Nash equilibrium (NE). Finally, we formalize the problem of entry
and spectrum sharing scheme selection for the entrant, and complete
our analysis by showing numerical results.


 The rest of this paper is organized as follows.
 Section  \uppercase\expandafter{\romannumeral2} reviews related
work, and Section  \uppercase\expandafter{\romannumeral3}
 describes the model.
In Sections \uppercase\expandafter{\romannumeral4} and
\uppercase\expandafter{\romannumeral5}, we consider a market with no
and with one incumbent, respectively, to study the user subscription
dynamics and revenue maximization. The problem of technology
selection is formalized in Section
\uppercase\expandafter{\romannumeral6}, and numerical results are
shown in Section \uppercase\expandafter{\romannumeral7}. Finally,
concluding remarks are offered in Section
\uppercase\expandafter{\romannumeral8}.

\section{Related Work}

Recently, communications markets have been attracting an
unprecedented amount of attention from various research communities,
due to their rapid expansion. For instance,
\cite{ShettyParekhWalrand} compared the profitability of different
spectrum sharing schemes in a femtocell market, whereas the user
subscription dynamics and the problem of spectrum sharing scheme
selection were neglected. In \cite{ShettySchwartzWalrand}, the
authors studied an Internet market and derived the optimal capacity
investments for Internet service providers under various regulation
policies. \cite{JinSenGuerinHosanagarZhang} studied technology
adoption and competition between incumbent and emerging network
technologies. Nevertheless,  only constant QoS functions were
considered in \cite{JinSenGuerinHosanagarZhang}. \cite{MussachioKim}
investigated market dynamics emerging when next-generation networks
and conventional networks coexist, by applying a market model that
consists of content providers, NSPs, and users. Nevertheless, the
level of QoS that a certain technology can provide was not
considered in the model. In \cite{HeWalrand}, the authors showed
that non-cooperative communications markets suffer from unfair
revenue distribution among NSPs and proposed a revenue-sharing
mechanism that requires cooperation among  NSPs. The behavior of
users and its impact on the revenue distribution, however, were not
explicitly considered in \cite{HeWalrand}. Without considering the
interplay between different NSPs, the authors in
\cite{HandeChiangCalderbankRangan} formulated a rate allocation
problem by incorporating the participation of content providers into
the model, and derived equilibrium prices and data rates. Another
paper related to our work is
\cite{ManshaeiFreudigerFelegyhaziMarbachHubaux} in which the authors
examined the evolution of network sizes in wireless social community
networks. A key assumption, based on which equilibrium was derived,
is that a social community network provides a higher QoS to each
user as the number of subscribers increases. While this assumption
is valid if network coverage is the dominant factor determining the
QoS or positive network externalities exist, it does not capture QoS
degradation due to, for instance, user traffic congestion incurred
at an NSP. \cite{AcemogluOzdaglar} focused on a communications
market with congestion costs and studied
 efficiency loss in terms of social welfare in both monopoly
and oligopoly markets. However, an implicit assumption in
\cite{AcemogluOzdaglar} is that users are homogeneous in the sense
that their valuations of QoS are the same. In
\cite{JiangParekhWalrand}, the authors considered users'
time-dependent utilities and proposed time-dependent pricing schemes
that maximize either social welfare or the service provider's
revenue. Although congestion effects were taken into account, the
convergence of the user subscription dynamics and competition
between NSPs were not studied.


\section{Model}

In this section, we provide a general model for
 the NSPs and users in a femtocell communications market.
Note that in addition to femtocell markets, our model also applies
to other communications markets such as cognitive radio markets.

\subsection{NSPs}
Consider a femtocell market with one incumbent NSP and a potential
entrant NSP, denoted by $\mathcal{S}_1$ and $\mathcal{S}_2$,
respectively. In the paper, we focus on the entry and technology
selection for the entrant. Thus, it is assumed that the incumbent
NSP $\mathcal{S}_1$ has already deployed its technology, whereas the
entrant NSP $\mathcal{S}_2$, if it decides to enter the market, can
select a technology from its $L$ available options, denoted by
$\{\mathcal{T}_1,\mathcal{T}_2,\cdots \mathcal{T}_L\}$, to maximize
its long-term profit. It should be pointed out that we use $L$
instead of a specific integer to keep the model general. For the
completeness of definition, we use $\mathcal{T}_0=\text{``Not
Enter''}$ to represent that the entrant
 chooses not to enter the
market, which yields zero long-term profit. In our considered
scenario, a technology represents a spectrum sharing scheme and the
available options for the entrant are ``split'', ``common'', and
``partially shared'' (or a subset of these three) in addition to
``Not Enter'', whereas the decision of spectrum allocation (i.e.,
how much bandwidth allocated to femtocells and macrocells) is not
explicitly considered in the paper. However, the spectrum allocation
decision can be captured if we treat different spectrum allocations
(but possibly with the same spectrum sharing scheme) as different
technologies in the set
$\{\mathcal{T}_0,\mathcal{T}_1,\mathcal{T}_2,\cdots
\mathcal{T}_L\}$. Moreover, we can also incorporate the entrant's
network capacity decision into our model by considering different
capacities (but possibly with the same spectrum sharing scheme
and/or spectrum allocation) as different technologies. Thus, a
technology can be considered as a combination of resources (e.g.,
spectrum) and the way to utilize available resources. Note that, as
we stated in Section I, technology selection is a long-term
decision, while users' subscription and NSPs' pricing decisions are
short-term and medium-term, respectively. Hence, the incumbent
cannot change its technology regardless of whether the entrant
enters the market or not \cite{ShettySchwartzWalrand}. We also
assume that the incumbent NSP $\mathcal{S}_1$ has sufficient
resources (e.g., network capacity) while the entrant NSP
$\mathcal{S}_2$ does not. An example that fits into our assumptions
on the NSPs is that the entrant is a small start-up. Hence, the QoS
provided by the incumbent degrades much more slowly than that
provided by the entrant and can be approximated as a constant
without losing much accuracy (see Fig.~\ref{qos_approx} and its
explanation for more details). On the other hand, as in
\cite{ChauWangChiu}\cite{AcemogluOzdaglarSrikant} where the authors
considered congestion effects because of limited resources, we
consider that the QoS provided by the entrant NSP $\mathcal{S}_2$
degrades with the number of its subscribers. Let $\lambda_i$ be the
fraction of users subscribing to NSP $\mathcal{S}_i$ for $i=1,2$.
Then $\lambda_1$ and $\lambda_2$ satisfy $\lambda_1,\lambda_2 \geq
0$ and $\lambda_1+\lambda_2\leq 1$. Also, let $q_i$ be the QoS
provided by NSP $\mathcal{S}_i$ for $i=1,2$. We assume that $q_1$ is
independent of $\lambda_1$ while $q_2$ is non-increasing in
$\lambda_2$. We use a function $g_{l}(\cdot)$ defined on $[0,1]$ to
express the QoS provided by NSP $\mathcal{S}_2$ as
$q_2=g_{l}(\lambda_2)$, if NSP $\mathcal{S}_2$ selects
$\mathcal{T}_l$ as its technology. We suppress the subscript $T_l$
when we analyze user subscription dynamics and NSP pricing decisions
 for notional convenience. Note that the QoS
metric can be anything that users care about (e.g., throughput,
delay, etc.). By considering average (normalized) throughput as the
QoS metric, we shall discuss in Section VII how to derive the QoS
function as a function of the number of subscribers. If the QoS
 is subject to multiple factors (e.g., throughput and
delay), then we can express the QoS as a multi-variable function
that takes into account all these factors.


\subsection{Users}

There
are a continuum of users that can potentially subscribe to one of the
NSPs for communication services. The continuum model approximates
well the real user population if there are a sufficiently large
number of users in the market so that each individual user is
negligible. As in
\cite{ManshaeiFreudigerFelegyhaziMarbachHubaux}\cite{ZemlianovVeciana},
we assume throughout this paper that each user can subscribe to at
most one NSP at any time instant.
Users are heterogeneous in the sense that they may value the same
level of QoS differently. Each user $k$ is characterized by a
non-negative real number $\alpha_k$, which represents its valuation
of QoS. Specifically, when user $k$ subscribes to NSP
$\mathcal{S}_i$, its utility is given by
\begin{equation}
\label{UtilityUserK} u_{k,i}= \alpha_k q_i - p_i,
\end{equation}
where $p_i$ is the subscription price charged by NSP
$\mathcal{S}_i$, for $i = 1,2$. Note that no other fees are charged
by the NSPs. Users that do not subscribe to either of the two NSPs
obtain zero utility. In \eqref{UtilityUserK},
the product of the QoS and the valuation of QoS represents  benefit
received by a user and the price represents  cost. We assume that a
user's utility is benefit from the service minus monetary cost. The
unit of user $k$'s valuation of QoS (i.e., $\alpha_k$) is chosen
such that $\alpha_k q_i$ has the same unit with that of the payment
$p_i$, for $i = 1, 2$. Note that in our model the NSPs are allowed
to engage in neither QoS discrimination nor price discrimination.
That is, all users subscribing to the same NSP receive the same QoS
and pay the same subscription price.

Now, we impose assumptions on the QoS function of NSP
$\mathcal{S}_2$, user subscription decisions, and the users'
valuations of QoS as follows.

\textit{Assumption 1:} For any technology
$\mathcal{T}_l$, $l=1,2,\cdots, L$, $g(\cdot)$ is a non-increasing and                                 
continuously differentiable\footnote{Since $g(\cdot)$ is defined on
$[0,1]$, we use a one-sided limit to define the derivative of
$g(\cdot)$ at $0$ and $1$, i.e.,
$g'(0)=\lim_{\lambda_2\to0^+}[g(\lambda_2)-g(0)]/(\lambda_2-0)$ and
$g'(1)=\lim_{\lambda_2\to1^-}[g(\lambda_2)-g(1)]/(\lambda_2-1)$.}
function, and $0< g(\lambda_2)< q_1$ for all $\lambda_2\in[0,1]$.

\textit{Assumption 2:}  Each user $k$ subscribes to NSP
$\mathcal{S}_i$ if $u_{k,i}>u_{k,j}$ and $u_{k,i}\geq0$ for
$i,j\in\{1,2\}$ and $i\not=j$. If $u_{k,1}=u_{k,2}\geq0$, user $k$
subscribes to NSP $\mathcal{S}_1$.\footnote{Specifying an
alternative tie-breaking rule (e.g., random selection between the
two NSPs) in case of $u_{k,1}=u_{k,2} \geq 0$ will not affect the
analysis of this paper, since the fraction of indifferent users is
zero under Assumption 3 and thus the revenue of the NSPs is
independent of
the tie-breaking rule. A similar remark
holds for the tie-breaking rule between
subscribing and not subscribing in case of $u_{k,i} = 0 \geq u_{k,j}$ for $i,j=\{1,2\}$ such that $i \neq j$.}

\textit{Assumption 3:} The users' valuations of QoS follow a
probability distribution whose probability density function (PDF)
$f(\cdot)$ is strictly positive and continuous on $[0,\beta]$ for
some $\beta > 0$. For completeness of definition, we have
$f(\alpha)=0$ for all $\alpha \notin [0,\beta]$. The cumulative
density function (CDF) is
given by $F(\alpha)=\int_{-\infty}^{\alpha}f(x)dx$ for all $\alpha \in \mathbb{R}$.

We briefly discuss the above three assumptions. Assumption~1
captures congestion effects that users experience when subscribing
to  NSP $\mathcal{S}_2$ with limited resources. Since NSP
$\mathcal{S}_1$ has more resources than NSP $\mathcal{S}_2$, it is
natural that $0< g(\lambda_2)< q_1$ for $\lambda_2\in[0,1]$ (see
Fig. \ref{qos_approx} for illustration). Assumption~2 can be
interpreted as a rational subscription decision. A rational user
will subscribe to the NSP that provides a higher utility if at least
one NSP provides a non-negative utility, and to neither NSP
otherwise. Assumption~3 can be considered as an expression of user
diversity in terms of the valuations of QoS. The lower bound on the
interval is set as zero to simplify the analysis, and as considered
in \cite{ShettyParekhWalrand}, this will be the case when there is
enough diversity in the users' valuations of QoS so that there are
non-subscribers for any positive price. 

\begin{figure}[t]
\begin{center}
\includegraphics[width=7.7cm]{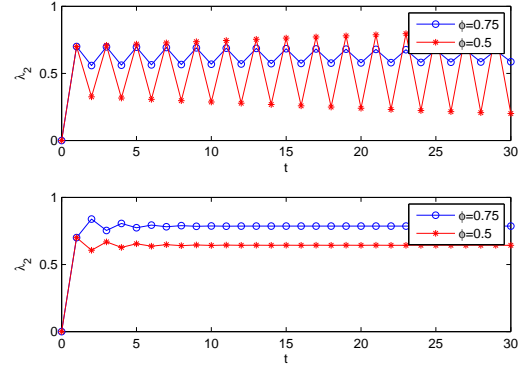}
\caption{User subscription dynamics in a market with no incumbent.
$u_{2,k}=\alpha_k\bar{q}_2-\delta \lambda_2 + \beta
\lambda_2^{\gamma} - p_2$. $p_2=0.3$, $\bar{q_2}=1.0$, $\delta=0.8$,
and $\gamma=0.2$. $\delta=1.2$ (upper) and $\delta=0.8$ (lower).}
\label{revise_mono_convergence_positive}
\end{center}
\end{figure}

\subsection{Assumptions and Remarks}

In the following, we
 list important assumptions made in this paper to further clarify our model and the scenario that we focus
 on.

 \textbf{a. Constant QoS provided by the incumbent:} The incumbent has
 sufficient (or over-provisioned) resources, such as available spectrum, and
 thus, it can provide a constant QoS to each user regardless of the number of
subscribers.

\textbf{b. No positive network externalities:} The QoS provided by
the entrant NSP is (weakly) decreasing in the number of its
subscribers.

\textbf{c. Lower QoS provided by the entrant than by the incumbent:}
The QoS provided by the entrant is lower than that provided by the
incumbent.

\textbf{d. Pricing-taking users:} Users take the prices set by the
NSPs as given, rather than anticipating the impacts of their
decisions on the prices.

\textbf{e. No switching cost:} There is no cost, referred to as
\emph{switching cost}, incurred when users change their subscription
decisions (see Section IV-A for more details).

Before proceeding with the analysis, we explain some of the
assumptions in the following remarks.

\textit{Remark 1 (Constant QoS provided by the
incumbent):} Given sufficient resources, the QoS provided by the
incumbent degrades sufficiently slowly such that it can be
approximated as a constant without losing much accuracy  (see Fig.
\ref{qos_approx} for more details). If the qualities of services
provided by both the incumbent and the entrant are degrading as more
users subscribe, then the price and the degradable QoS  will jointly
affect the users' subscription decisions.  Although the
corresponding quantitative results will be different, the
qualitative results remain unchanged. For example, the entrant NSP
providing a uniformly lower QoS needs to charge a lower price in
order to maximize its revenue.

\textit{Remark 2 (No positive network externalities):}  We note that
compared to positive network externalities,\footnote{A primary
source of positive network externalities is the (supply-side)
economies of scale. Specifically, as there are more subscribers in a
femtocell network, the NSP can make an investment to improve the
service quality.} negative network externalities are typically
considered as dominating effects in wireless networking research,
including femtocell research (e.g., \cite{ShettyParekhWalrand}). For
instance, when more users subscribe, congestion and interferences
become intolerable, if the network resource (e.g., capacity) is not
sufficient, and will significantly affect the users' experiences. In
general, suppressing the positive network externalities while only
focusing on negative network effects (i.e., congestion effects) is a
common approach in wireless networking and some operational research
to studying the interaction among multiple network service providers
(see, e.g.,
\cite{ShettyParekhWalrand}\cite{ShettySchwartzWalrand}\cite{AcemogluOzdaglar}\cite{JohariWeintraubVanRoy}\cite{ZemlianovVeciana}\cite{ChauWangChiu}\cite{KunNiyatoWang}
and references therein). If positive network externalities are also
taken into account in our model, there may exist multiple and
possibly unstable
 equilibrium points in the user subscription dynamics.
By considering the utility function
$u_{i,k}=\alpha_k\bar{q}_i-\delta \lambda_i + \phi
\lambda_i^{\gamma} - p_i$ for user $k$ where $0\leq\phi<\gamma$,
$\gamma<1$, and $\phi \lambda_i^{\gamma}$ captures the positive
network externalities), we show in Fig.
\ref{revise_mono_convergence_positive} the user subscription
dynamics with positive network externalities in a market with no
incumbent. The details of specifying the user subscription dynamics
are provided in Section IV-A. We see that if $\delta=0.8$  (i.e.,
not too large compared to $\bar{q}_2=1.0$, or the effects of
negative externalities do not increase significantly when more users
subscribe), then the convergence can be observed (from different
starting points). Fig. \ref{revise_mono_convergence_positive} only
shows a few instances for the ease of illustration, while more
numerical results can be shown to support our statement.  Note that,
because of the term $\phi \lambda_i^{\gamma}$ with $\gamma<1$
representing the positive network externalities, we cannot
theoretically guarantee the convergence starting from any initial
points and given any charged price. Nevertheless, with
$\gamma\geq1$, it can be shown based on contraction mapping
\cite{BertsekasTsitsiklis} that the existence of a unique
equilibrium point and the convergence can be guaranteed for any
charged price and any initial point $\lambda_2^0\in[0,1]$ if the
condition
\begin{equation}
\label{ConvergenceConditionPositive}
\left[{\max_{\alpha\in[0,\beta]}f(\alpha)}\right]\cdot\frac{\phi\gamma+\delta}{\bar{q}_2}<1
\end{equation}
is satisfied.  We see from \eqref{ConvergenceConditionPositive} that
$\delta$ cannot be too large given $\phi$ and $\gamma$. This is
similar to our derived sufficient condition for the convergence of
user subscription dynamics without positive network externalities.
Similar results hold for the market with one incumbent, and are not
shown here for brevity. A comprehensive investigation of the
coexistence of both positive and negative network externalities will
be left for our future work.

 \textit{Remark 3 (Lower QoS provided by the
entrant than by the incumbent):} Given the three-stage decisions
shown in Fig. \ref{timing}, we implicitly assume that the NSP can
afford any available technology at the beginning and hence, there is
no need to ``upgrade'' the initially chosen technology afterwards.
In general, there are two types of constraints -- budget and
technology\footnote{Recall that a technology can be considered as a
combination of resources (e.g., spectrum) and the way to utilize
available resources.}
 -- that limit the entrant's
technology selection. ``No budget constraint'' is assumed in the
sense that the entrant can choose any  technologies which are
\emph{available} for its selection. Thus, technology selection is
primarily subject to the technology availability. Specifically, we
focus on the case in which the technology available for the entrant
is inferior to the incumbent's in terms of QoS provisioning (i.e.,
$g(\lambda_2)<q_1$ for $\lambda_2\in[0,1]$). In particular,  in the
case where the entrant has fewer resources than the incumbent (but
the way to utilize the resources, e.g., spectrum sharing scheme, is
the same), the QoS offered by the entrant will be lower than that
offered by the incumbent (see, e.g., Fig.~\ref{qos_approx}). This
situation may arise in several practical scenarios. For example, if
the incumbent is a wireless operator serving primary users while the
incumbent operates a cognitive radio network serving secondary users
that only \emph{opportunistically} access to the ``spectrum holes''.
Another scenario is that upon the entrant's entry into a femtocell
market,
 only very limited spectrum is
available, while the incumbent has already obtained a much larger
range of spectrum. In each of these scenarios, we expect that the
QoS provided by the entrant is not as good as the incumbent's, even
though the incumbent's budget is sufficient to cover its entry and
technology selection.

\textit{Remark 4 (Price-taking users):} The assumption of
price-taking users is reasonable when there are a sufficiently large
number potential subscribers. In such cases, the impact of a single
individual's subscription decision on the decisions of the NSPs is
negligible. In this paper, we use a continuum model to analyze the
case of a sufficiently large number potential subscribers.

\textit{Remark 5 (Applicability of our model):} Besides the
femtocell market we focus on, our proposed model applies to a number
of other communications markets. In particular, we can apply
 the model to study the spectrum acquisition decision (i.e., how
 much spectrum to purchase/lease from the spectrum owner) made by
 a small wireless carrier providing wireless cellular services, by
 a mobile virtual network operator (MVNO) \cite{CheboldaeffMVNO} or
 by an entrant
providing cognitive radio access services \cite{HaykinCognitive}. In
such scenarios,  the long-term ``technology selection'' in our model
becomes ``spectrum acquisition decision'', whereas the medium-term
pricing decision and short-term user subscription decisions remain
unaffected.


\section{Femtocell Market With No Incumbent}

In this section, we study user subscription
dynamics and revenue maximization for the entrant in a
femtocell  market with no incumbent. In this scenario, the
entrant becomes the monopolist in the market. In practice, this
corresponds to an emerging market which an entrant tries to explore.
We study first the user subscription dynamics and then the problem
of revenue maximization, based on which the entrant can finally
select its technology with which its profit is maximized.

\subsection{User Subscription Dynamics}

When the entrant NSP $\mathcal{S}_2$ operates in a market with no
incumbent, each user has a choice of whether to subscribe to NSP
$\mathcal{S}_2$ or not at each time instant. Since the QoS provided
NSP $\mathcal{S}_2$ is varying with the fraction of its
subscribers,\footnote{``Fraction of subscribers'' of an NSP is used
throughout this paper to mean the proportion of users in the market
that subscribe to this NSP. } each user will form a belief on the
QoS of NSP $\mathcal{S}_2$ when it makes a subscription decision. To
describe the dynamics of user subscription, we construct and analyze
a dynamic model which specifies how users form their beliefs and
make decisions based on their beliefs. We consider a discrete-time
model with time periods indexed $t = 1,2,\ldots$. At each period
$t$, user $k$ holds a belief on the QoS of NSP $\mathcal{S}_2$,
denoted by $\tilde{q}_{2,k}^t$  and the subscript $k$ denotes the
user index, and makes a subscription decision to maximize its
expected utility in the current period.\footnote{An example
consistent with our subscription timing is a ``Pay-As-You-Go'' plan
in which a subscribing user pays a fixed service charge for a period
of time (day, week, or month) and is free to quit its subscription
at any time period, effective from the next time period.} Then, user
$k$ subscribes to NSP $\mathcal{S}_2$ at period $t$ if and only if
$\alpha_k \tilde{q}_{2,k}^t \geq p_2$. As in
\cite{JinSenGuerinHosanagarZhang}\cite{ManshaeiFreudigerFelegyhaziMarbachHubaux},
an implicit assumption is that that, other than the subscription
price, there is no cost involved in subscription decisions (e.g.,
initiation fees, termination fees, device prices). We specify that
every user expects that the QoS in the current period is equal to
that in the previous period. That is, we have $\tilde{q}_{2,k}^t =
g(\lambda_2^{t-1})$ for $t = 1,2,\ldots$, where $\lambda_2^t$ is the
fraction
of subscribers at period $t$.\footnote{%
This model of belief formation is called naive or static
expectations in \cite{EvansHonkapohja}. A similar dynamic model of
belief formation and decision making has been extensively adopted in
the existing literature (see, e.g.,
\cite{JinSenGuerinHosanagarZhang}\cite{ManshaeiFreudigerFelegyhaziMarbachHubaux}\cite{ZemlianovVeciana}).}

By substituting
$\tilde{q}_{2,k}^t=g(\lambda_2^{t-1})$ into $\alpha_k
\tilde{q}_{2,k}^t \geq p_2$, we can see that user $k$ subscribes to
NSP $\mathcal{S}_2$ if and only if $\alpha_k\geq
\frac{p_2}{g(\lambda_2^{t-1})}$.  That is, only those users with a
valuation of QoS greater than or equal to
$\frac{p_2}{g(\lambda_2^{t-1})}$ will subscribe to NSP
$\mathcal{S}_2$ at time $t$. Thus, the fraction of subscribers of
NSP $\mathcal{S}_2$ evolves following a sequence
$\{\lambda_2^t\}_{t=0}^{\infty}$ in $[0,1]$ generated by
\begin{equation}
\label{EvolutionProcess2}
\lambda_2^t=h_m(\lambda_2^{t-1}) \triangleq 1-F\left(\frac{p_2}{g(\lambda_2^{t-1})}\right),
\end{equation}
for $t=1,2,\ldots$, starting from a given initial point $\lambda_2^0\in[0,1]$. Note that the price $p_2$ of NSP $\mathcal{S}_2$
is held fixed over time.
Given the user subscription dynamics \eqref{EvolutionProcess2}, we are interested in
whether the fraction of subscribers will stabilize in the long run and, if so,
to what value.
As a first step, we define an equilibrium point of the user subscription dynamics.

\textit{Definition 1:} $\lambda_2^*$ is an \emph{equilibrium} point
of the user subscription dynamics in the monopoly market of NSP
$\mathcal{S}_2$ if it satisfies
\begin{equation}
\label{EquilibriumEvolutionProcess}
h_m(\lambda_2^*) = \lambda_2^*.
\end{equation}

Definition \ref{EquilibriumEvolutionProcess} implies that once an
equilibrium point is reached, the fraction of subscribers remains
the same from that point on. Thus, equilibrium points are natural
candidates for the long-run fraction of subscribers. The following
Proposition, whose proof is deferred to
to Appendix~A, establishes the
existence and uniqueness of an equilibrium point.

\begin{prop}
\label{ExistenceOfEquilibriumPointMonopoly} For any non-negative
price $p_2$, there exists a unique equilibrium point of the user
subscription dynamics in the market of NSP $\mathcal{S}_2$.
\hfill$\square$
\end{prop}

Although the analysis in this paper applies to a
general QoS function $g(\lambda_2)$, we consider a class of simple
QoS functions defined below in order to obtain a closed-form
expression of the equilibrium point and solve the revenue
maximization problem explicitly.

\textit{Definition 2:} The QoS function $g(\cdot)$ is
\emph{linearly-degrading} if $g(\lambda_2)=\bar{q}_2- c \lambda_2$
for all $\lambda_2 \in [0,1]$, for some $\bar{q}_2>0$ and $c\in[0,\bar{q}_2)$. In
particular, a linearly-degrading QoS function with $c=0$, i.e.,
$g(\lambda_2)=\bar{q}_2$ for all $\lambda_2 \in [0,1]$, is referred to
as a constant QoS function.

Linearly-degrading QoS functions model a variety of applications
including flow control in \cite{ZhangDouligeris} and capacity
sharing in \cite{ChauWangChiu}. More
importantly, it can be viewed as an affine approximation of real QoS
functions and we shall see in the numerical results that the affine
approximation is reasonably close to the actual QoS functions.
With a linearly-degrading QoS function and uniformly distributed
valuations of QoS, we can obtain a simple closed-form expression of
the equilibrium point. Specifically, with $g(\lambda_2)=\bar{q}_2-c
\lambda_2$ for $\lambda_2\in[0,1]$ and $f(\alpha)=1/\beta$ for
$\alpha\in[0,\beta]$, the equilibrium point of the user subscription
dynamics in the market of NSP $\mathcal{S}_2$ can be expressed as a
function of $p_2$ as follows:
\begin{equation}
\label{EquilibriumSolutionUniformLinear}
\lambda_2^*(p_2) =\left\{
\begin{matrix}
\frac{\bar{q}_2+c-\sqrt{(\bar{q}_2-c)^2 + \frac{4cp_2}{\beta}}}{2c},&&\hfill\text{for }p_2\in[0,\beta \bar{q}_2],\\
0,\hfill&&\hfill\text{for }p_2\in(\beta\bar{q}_2,\infty),
\end{matrix}\right.
\end{equation}
if $c \in (0, \bar{q}_2)$, and $\lambda_2^*(p_2)=\max\{0,
1-p_2/(\beta \bar{q}_2)\}$ if $c=0$.

Our equilibrium analysis so far guarantees the existence of a unique
stable point of the user subscription dynamics. However, it does not
discuss whether the unique stable point will be eventually reached.
To answer this question, we turn to the analysis of the convergence
properties of the user subscription dynamics. The convergence of the
user subscription dynamics is not always guaranteed, especially when
the QoS provided by the monopolist degrades rapidly with respect to
the fraction of subscribers. As a hypothetical example, suppose that
only a small fraction of users subscribe to NSP $\mathcal{S}_2$ at
period $t$ and each subscriber obtains a high QoS. In our model of
belief formation, users expect that the QoS will remain high at
period $t+1$, and thus a large fraction of users subscribe at period
$t+1$, which will result in a low QoS at period $t+1$. This in turn
will induce a small fraction of subscribers at period $t+2$. When
the QoS is very sensitive to the fraction of subscribers, the user
subscription dynamics  may oscillate around or diverge away from the
equilibrium point and thus convergence may not be obtained. The
following theorem provides a sufficient condition under which the
user subscription dynamics always converges.

\begin{theorem}
\label{ConvergenceOfEvolution} For any non-negative price $p_2$, the user
subscription dynamics specified by (\ref{EvolutionProcess2}) converges to
the unique equilibrium point starting from any initial point $\lambda_2^0 \in [0,1]$ if
\begin{equation}
\label{ConvergenceCondition}
\max_{\lambda_2\in[0,1]}\left\{-\frac{g'(\lambda_2)}{g(\lambda_2)}\right\}<\frac{1}{K},      
\end{equation}
where $K=\max_{\alpha\in[0,\beta]}f(\alpha)\alpha$.                                    
\end{theorem}
{\proof See Appendix~B. \hfill$\square$}

By applying Theorem~1 to linearly-degrading QoS functions, we obtain
the following result.

\begin{corollary}\label{CorollaryLinearMonopoly}
If the QoS function $g(\cdot)$ is linearly-degrading, i.e., $g(\lambda_2)=\bar{q}_2-c \lambda_2$
for $\lambda_2\in[0,1]$, and
\begin{equation}
\label{ConvergenceConditionLinear} \frac{c}{\bar{q}_2}<\frac{1}{1+K}
\end{equation}
where $K=\max_{\alpha\in[0,\beta]}f(\alpha)\alpha$, then the user
subscription dynamics converges to the unique equilibrium point
starting from any initial point $\lambda_2^0 \in [0,1]$.
\hfill$\square$
\end{corollary}

The condition (\ref{ConvergenceCondition}) in
Theorem~\ref{ConvergenceOfEvolution} is sufficient but not necessary
for the convergence of the user subscription dynamics. In
particular, we observe through numerical simulations that in some
cases (e.g., $g(\lambda_2)=1-0.9\lambda_2$ for $\lambda_2\in[0,1]$
and $f(\alpha) = 1$ for $\alpha \in [0,1]$) the user subscription
dynamics converges for a wide range of prices although the condition
(\ref{ConvergenceCondition}) is violated. Nevertheless, the
sufficient condition provides us with the insight that if QoS
degradation is too fast (i.e., $- K g'(\lambda_2)$ is larger than
$g(\lambda_2)$ for some $\lambda_2 \in [0,1]$), the dynamics may
oscillate or diverge. If our analysis is applied to study
the spectrum acquisition decision, a practical implication
of the derived convergence condition (which may also
be mapped into the spectrum requirement) is that the acquired spectrum
should be sufficiently large such that the congestion does
not grow too rapidly when more users subscribe \cite{RenVanderschaarTechnicalReport}.

We set up our basic model by assuming that all the users will
simultaneously make their subscription decisions at the beginning of
each decision period, i.e., ``simultaneous/synchronous move''. However, it should be
noted that we can generalize the user subscription dynamics by
assuming that only $\epsilon$ fraction of users, where $\epsilon \in
(0,1]$, change their subscription decisions in each period. In this
generalized scenario, not all the users make their decisions
simultaneously, while only $\epsilon$ fraction of users in the
market do. We still assume that all the users that change
their subscription decisions expect that the QoS they
receive in the next time period will be the same as that
in the current time period.
This can be viewed as ``asynchronous move'', under
which the user subscription dynamics is generated by
\begin{equation}
\label{EvolutionProcess3} \lambda_2^t=(1-\epsilon) \lambda_2^{t-1} +
\epsilon h_m(\lambda_2^{t-1})
\end{equation}
for $t=1,2,\ldots$, starting from an initial point
$\lambda_2^0\in[0,1]$.  For the more general user subscription dynamics in
\ref{EvolutionProcess3}, our original existence and uniqueness
analysis is still valid, whereas the convergence analysis
(Theorem~1) is affected and the sufficient convergence condition
 is modified as
\begin{equation}
\label{ConvergenceCondition2}
\max_{\lambda_2\in[0,1]}\left\{-\frac{g'(\lambda_2)}{g(\lambda_2)}\right\}<\frac{1}{\epsilon K}.      
\end{equation}
As the condition (\ref{ConvergenceCondition2}) is more easily
satisfied for a smaller $\epsilon$, we see that there is a trade-off
between the guarantee of convergence and the speed of convergence.
There exist other dynamics, such as continuous-time dynamics, modeling
the user subscription process and interested readers may refer
to \cite{JinSenGuerinHosanagarZhang} for a detailed analysis.



\begin{figure}[t]
\begin{center}
\includegraphics[width=7.7cm]{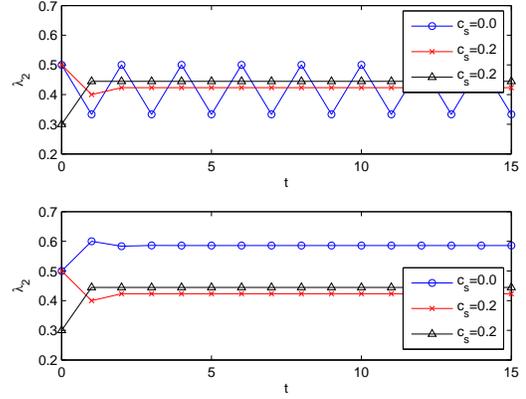}
\caption{User subscription dynamics in a market with no incumbent.
$p_2=0.4$. For $\lambda_2\in[0,1]$, $g(\lambda_2)=1.2-1.2\lambda_2$
(upper) and $g(\lambda_2)=1.2-0.4\lambda_2$ (lower).}
\label{revise_mono_convergence_switchingcost}
\end{center}
\end{figure}

Next, we  discuss the cost involved when users change their subscription decisions.
For simplicity, we assume  that the costs of activating and terminating the subscription are the same,
and we refer to this cost
as \emph{switching} cost denoted by $c_s$,
 which includes, but is not
limited to, time spent
 in calling the customer service, activation fees and early
termination fees. By charging this cost, the NSP creates the effect
of user ``lock-in'', which we note may result in multiple
equilibrium points and different convergence
 behaviors, subject to the initial point.
For instance, in the extreme case in which the cost is so high
 (e.g., greater than $\beta g(0)$ which is the highest benefit that
 a user can possibly gain by subscribing to NSP $\mathcal{S}_2$) that
no users would like to change their subscription decisions, every
possible value of $\lambda_2\in[0,1]$ is an equilibrium point.
In general, if user $k$ is a subscriber in the time period $t$, it will continue
the subscription in the next time period $t+1$ if
\begin{equation}
\begin{split}
\label{SubscriptionDecisionSwitchingCost} \alpha_k g(\lambda_2^t) -
p_2  \geq -c_s.
\end{split}
\end{equation}
On the other hand, if user $k$ is not a subscriber in the time
period $t$, it will choose to subscribe to the NSP in the next time
period $t+1$ if
\begin{equation}
\begin{split}
\label{SubscriptionDecisionSwitchingCost_2} \alpha_k g(\lambda_2^t)
- p_2 -c_s\geq 0.
\end{split}
\end{equation}
While rigorous analysis of $c_s$ is left as our future work, we show
in Fig. \ref{revise_mono_convergence_switchingcost} the impact of
switching cost $c_s$ on the users' subscription decisions. The upper
plot indicates that switching costs may make the user subscription
dynamics converge even though the QoS degrades rapidly such that the
user subscription dynamics does not converge without switching
costs. We explain this point by noting that, with switching costs,
fewer users will change their subscription decisions and hence the
user subscription dynamics converges under milder conditions. With
switching costs imposed, it can also be seen from  Fig.
\ref{revise_mono_convergence_switchingcost} that there may exist
multiple equilibria in the user subscription dynamics and the
equilibrium, to which the user subscription dynamics converges,
depends on the initial point. Since the analysis of the NSP's
pricing decision and technology selection largely relies on the
equilibrium point of the user subscription dynamics,  the existence
of multiple equilibrium points is challenging to deal with and loses
mathematical tractability. Thus, as in the existing related
literature (e.g.,
\cite{JinSenGuerinHosanagarZhang}\cite{ManshaeiFreudigerFelegyhaziMarbachHubaux}\cite{ChauWangChiu}),
the switching cost is not considered in our paper. Moreover,
neglecting the switching cost is particularly applicable in a
setting where handover and service provider selection in real time
are possible.

\subsection{Revenue Maximization} \label{sec:RevenueMonopoly}

Building on the equilibrium analysis of the user subscription dynamics, we are
now interested in finding an optimal price of NSP $\mathcal{S}_2$
that maximizes its equilibrium revenue in the market with no incumbent.\footnote{%
By focusing on equilibrium revenue, we implicitly assume that the
unique equilibrium point is reached within a relatively short period
of time.} Note that the optimal revenue is associated with the
technology selected by NSP $\mathcal{S}_2$. To keep the notion
succinct, we omit the subscript of $\mathcal{T}_l$ in the revenue
and express it as
\begin{equation}
\label{RevenueNSPMonopolyS2} R_2(p_2)=
p_2\lambda_2^*(p_2),
\end{equation}
where $\lambda_2^*(p_2)$ is the equilibrium point of the user
subscription dynamics at price $p_2$. It can be shown that
$\lambda_2^*(0) = 1$, $\lambda_2^*(\cdot)$ is strictly decreasing on
$[0, \beta g(0)]$, and $\lambda_2^*(p_2)=0$ for all $p_2 \geq \beta
g(0)$, where $\beta$ is the maximum valuation of QoS of all the
users. As a result, NSP $\mathcal{S}_2$ will gain a positive revenue
only if it sets a price $p_2$ in $(0, \beta g(0))$, and thus a
revenue-maximizing price lies in $(0, \beta g(0))$. However, it is
difficult to directly obtain an explicit expression of $p_2\in(0,
\beta g(0))$ that maximizes $R_2(p_2)$ even when the QoS function is
linearly-degrading and the users' valuations of QoS are uniformly
distributed, since $\lambda_2^*(p_2)$ is a complicated function of
$p_2$ as can be seen in \eqref{EquilibriumSolutionUniformLinear}. In
the following analysis, we reformulate the revenue maximization
problem by applying the marginal user principle\footnote{In the
market with no incumbent, marginal users are users that are
indifferent between subscribing and not subscribing to NSP
$\mathcal{S}_2$ given the received QoS and the charged price. In our
model, a marginal user receives zero utility.}
\cite{ShakkottaiSrikantOzdaglarAcemoglu}\cite{AcemogluOzdaglarSrikant}.
Specifically, we change the choice variable in the revenue
maximization problem.

Suppose that a marginal user exists, whose valuation of QoS is
denoted by $\alpha$. Then from the utility function in
(\ref{UtilityUserK}), we can see that all the users with valuations
of QoS greater than $\alpha$ receive a positive utility and thus
subscribe to NSP $\mathcal{S}_2$
\cite{ShettyParekhWalrand}\cite{ShakkottaiSrikantOzdaglarAcemoglu}.
Hence, when a marginal user has a valuation of QoS $\alpha \in
[0,\beta]$, the fraction of subscribers  is given by $\lambda_2 = 1
- F(\alpha)$. Also, for a given price $p_2 \in [0, \beta g(0)]$,
there exists a unique valuation of QoS of a marginal user $\alpha
\in [0,\beta]$, and the relationship between $p_2$ and $\alpha$ is
given by
\begin{equation}
\label{PriceAlpha} p_2 = \alpha g(1-F(\alpha)).
\end{equation}
 Based on the above relationships between $p_2$, $\alpha$, and $\lambda_2$, we can formulate the
revenue maximization problem using different choice variables as
follows:
\begin{equation}
\begin{split}
\label{RevenueNSPMonopolyS2Equivalent} \max_{p_2 \in [0, \beta
g(0)]} p_2 \lambda_2^*(p_2) &= \max_{\alpha\in[0,\beta]}\alpha
g(1-F(\alpha)) \left[1-F(\alpha)\right]\\
&= \max_{\lambda_2\in[0,1]} F^{-1}(1-\lambda_2) g(\lambda_2)
\lambda_2,
\end{split}
\end{equation}
where $F^{-1}(\cdot)$ is the inverse function of $F(\cdot)$ defined on $[0,1]$.\footnote{%
We define $F^{-1}(0)=0$ and $F^{-1}(1)=\beta$.} It is clear that a
solution to each of the above three problems exists, since the
constraint set is compact and the objective function is continuous.
Let $p^*$, $\alpha^*$, and $\lambda_2^{**}$ be a solution to each
respective problem in \eqref{RevenueNSPMonopolyS2Equivalent}. By
imposing an assumption on the distribution of the users' valuations
of QoS, we obtain upper and lower bounds on $p^*$, $\alpha^*$, and
$\lambda_2^{**}$ in Proposition~\ref{UpperLowerBoundsMonopoly},
whose proof is given 
in Appendix~C.

\begin{prop}
\label{UpperLowerBoundsMonopoly} Suppose that $f(\cdot)$ is non-increasing
on $[0,\beta]$. Then optimal variables solving the revenue maximization problem
in (\ref{RevenueNSPMonopolyS2Equivalent}) satisfy
$F^{-1}(1/2) g(1/2)\leq p_2^*<\beta g(0)$,
$F^{-1}(1/2)\leq\alpha^*<\beta$, and $0<\lambda_2^{**}\leq 1/2$.\hfill$\square$
\end{prop}

The non-increasing property of $f(\cdot)$ can be considered as
representing a class of emerging markets where there are fewer users
with higher valuations of QoS provided by the NSP \cite{Tirole}.
Proposition \ref{UpperLowerBoundsMonopoly} shows that when the
monopolist maximizes its revenue in an emerging market, no more than
a half of the users, only those whose valuations are sufficiently
high, are served. In other words, in an emerging market, the NSP
will serve a minority of users with high valuations to maximize its
revenue. Since a uniform distribution satisfies the non-increasing
property, applying Proposition \ref{UpperLowerBoundsMonopoly} to the
case of a uniform distribution of the users' valuations of QoS
(i.e., $f(\alpha)=1/\beta$ and $F(\alpha) = \alpha/\beta$ for
$\alpha\in[0,\beta]$) yields $(\beta/2) g(1/2)\leq p_2^*<\beta g(0)$
and $\beta/2\leq\alpha^*<\beta$. If, in addition, the QoS function
satisfies the sufficient condition (\ref{ConvergenceCondition}) for
convergence, we obtain tighter bounds on optimal variables.

\begin{corollary}
\label{UpperLowerBoundsMonopolyMoreSpecific} Suppose that
$f(\alpha)=1/\beta$ for $\alpha\in[0,\beta]$ and
$- g'(\lambda_2) / g(\lambda_2) < 1$ for all $\lambda_2 \in [0,1]$.
Then optimal variables solving the revenue maximization problem
in (\ref{RevenueNSPMonopolyS2Equivalent}) satisfy
$(\beta/2) g(1/2)\leq p_2^*<[(\sqrt{5}-1)/2] \beta g((3-\sqrt{5})/2)$,
$\beta/2 \leq\alpha^*<[(\sqrt{5}-1)/2] \beta$, and $(3-\sqrt{5})/2<\lambda_2^{**}\leq 1/2$.
\end{corollary}
\proof The proof is omitted due to space limitations.
\hfill$\square$

 With a uniform distribution of the users'
valuations of QoS and a linearly-degrading QoS function, we can
obtain explicit expressions of optimal variables of the revenue
maximization problem as follows:
\begin{eqnarray}
\label{OptimalAlpha}
&&\alpha^*=\frac{2c-\bar{q}_2+\sqrt{\bar{q}_2^2+c^2-c\bar{q}_2}}{3c}\beta,\\
\label{OptimalLamda}
&&\lambda_2^{**}=\frac{c+\bar{q}_2-\sqrt{\bar{q}_2^2+c^2-c\bar{q}_2}}{3c},
\end{eqnarray}
and $p_2^*=\alpha^*(\bar{q}_2-c\lambda_2^{**})$. The high-level
insight from this result is that the optimal price maximizing the
NSP's revenue should be decreased if the QoS degrades more quickly
and that the optimal market share is independent of the interval on
which the users' valuation of QoS is uniformly distributed.
Moreover, the result quantifies the impacts of the QoS function
(i.e., maximum QoS and degrading rate) as well as the users'
valuation of QoS on the optimal price. For instance, as shown in
Fig. \ref{revise_mono_revenue_lambda_2}, the NSP does not incur a
significant revenue loss if its equilibrium market share is around
one half, whereas its revenue loss is nearly 10\% and more if its
equilibrium market share is less than 0.4 or greater than 0.6. This
indicates that the NSP's revenue is near to its optimum if the NSP
serves around one half of the market, while both under-serving and
over-serving will significantly reduce the NSP's revenue. Due to the
implicit and explicit coupling involved in our considered
three-decision making process, it is difficult to see how the
quantitative results in the pricing decision stage directly affects
the entrant's long-term technology selection. Nevertheless, solving
the revenue maximization problem (i.e., medium-term problem)
explicitly serves as a basis for the entrant to decide whether or
not to enter the market and select the technology that maximizes its
long-term profit (i.e., long-term problem).

Finally, we note that in order to maximize its equilibrium revenue,
the entrant needs to know the distribution of the users' valuations
of QoS  by conducting market surveys and using data-mining and
learning techniques. The details of information acquisition are
beyond the scope of this paper.


\section{Femtocell Market With One Incumbent}

In this section, we analyze user subscription dynamics and market
competition for the entrant in a femtocel market with one
incumbent. In other words, the two NSPs operate and compete against
each other in a duopoly market.

\subsection{User Subscription Dynamics}

With the two NSPs operating in the market, each user has three
possible choices at each time instant: subscribe to NSP
$\mathcal{S}_1$, subscribe to NSP $\mathcal{S}_2$, and subscribe to
neither.  As in the market with no incumbent, we consider a dynamic
model in which the users update their beliefs and make subscription
decisions at discrete time period $t = 1,2,\ldots$. The users expect
that the QoS provided by NSP $\mathcal{S}_2$ in the current period
is equal to that in the previous period and make their subscription
decisions to maximize their expected utility in the current period
\cite{ManshaeiFreudigerFelegyhaziMarbachHubaux}. We assume that,
other than the subscription price, there is no cost involved in
subscription decisions (e.g., initiation fees, termination fees)
when users switch between NSP $\mathcal{S}_1$ and NSP
$\mathcal{S}_2$ \cite{JinSenGuerinHosanagarZhang}. By Assumption 2,
at period $t=1,2\cdots$, user $k$ subscribes to NSP $\mathcal{S}_1$
if and only if
\begin{eqnarray}
\label{ConditionSubscribeS1First}
\alpha_k q_1  -p_1 \geq \alpha_k g(\lambda_2^{t-1}) - p_2 \text{ and }
\alpha_k q_1  -p_1 \geq 0,
\end{eqnarray}
to NSP $\mathcal{S}_2$ if and only if
\begin{eqnarray}
\label{ConditionSubscribeS2First}
\alpha_k g(\lambda_2^{t-1}) - p_2 > \alpha_k q_1  -p_1 \text{ and }
\alpha_k g(\lambda_2^{t-1}) - p_2 \geq 0,
\end{eqnarray}
and to neither NSP if and only if
\begin{eqnarray}
\label{ConditionSubscribeNeither}
\alpha_k q_1  -p_1 < 0 \text{ and }
\alpha_k g(\lambda_2^{t-1}) - p_2 < 0.
\end{eqnarray}

By solving
\eqref{ConditionSubscribeS1First}--\eqref{ConditionSubscribeNeither},
it can be shown that, given the prices $(p_1,p_2)$, the user
subscription dynamics is described by a sequence
$\{(\lambda_1^t,\lambda_2^t)\}_{t=0}^{\infty}$ in $\Lambda =
\{(\lambda_1,\lambda_2)\in\mathbb{R}_+^2\;|\;\lambda_1+\lambda_2\leq1\}$
generated by
\begin{align}
\label{EvolutionDuopolyS1}
\lambda_1^t&=h_{d,1}(\lambda_1^{t-1},\lambda_2^{t-1}) \triangleq 1-F\left(\frac{p_1-p_2}{q_1-g(\lambda_2^{t-1}) }\right),\\
\label{EvolutionDuopolyS2}
\lambda_2^t &=h_{d,2}(\lambda_1^{t-1},\lambda_2^{t-1}) \triangleq
F\left(\frac{p_1-p_2}{q_1-g(\lambda_2^{t-1}) }\right) -
F\left(\frac{p_2}{g(\lambda_2^{t-1})}\right)
\end{align}
if $p_1/q_1>p_2/g(\lambda_2^{t-1})$, and by
\begin{align}
 \label{EvolutionDuopolyS1_2}
\lambda_1^t&=h_{d,1}(\lambda_1^{t-1},\lambda_2^{t-1}) \triangleq 1 - F\left(\frac{p_1}{q_1}\right),\\
 \label{EvolutionDuopolyS2_2}
\lambda_2^t &= h_{d,2}(\lambda_1^{t-1},\lambda_2^{t-1}) \triangleq 0
 \end{align}
if $p_1/q_1 \leq p_2/g(\lambda_2^{t-1})$, for $t = 1,2,\ldots$,
starting from a given initial point $(\lambda_1^0, \lambda_2^0)$.
Note that there are two regimes of the user subscription dynamics,
and which regime governs the dynamics depends on the relative values
of the \emph{prices per QoS}, i.e., $p_1/q_1$ and
$p_2/g(\lambda_2^{t-1})$. Specifically, if the
price per QoS offered by NSP $\mathcal{S}_1$ is higher than that
offered by NSP $\mathcal{S}_2$ (i.e.,
$p_1/q_1>p_2/g(\lambda_2^{t-1})$), then users who are sensitive to
prices (i.e., those whose valuations are not sufficiently high and
lie between $\frac{p_1-p_2}{q_1-g(\lambda_2^{t-1})}$ and
$\frac{p_2}{g(\lambda_2^{t-1})}$) will prefer NSP $\mathcal{S}_2$ to
NSP $\mathcal{S}_1$.

We give the definition of an equilibrium point, which is similar to
Definition 1.

\textit{Definition 3:} $(\lambda_1^*, \lambda_2^*)$ is an
\emph{equilibrium} point of the user subscription dynamics in
duopoly the
 market of NSP $\mathcal{S}_1$ and $\mathcal{S}_2$ if it satisfies
\begin{equation}
 \label{EquilibriumEvolutionProcessDuopoly}
h_{d,1}(\lambda_1^*,\lambda_2^*) = \lambda_1^* \text{  and  }
h_{d,2}(\lambda_1^*,\lambda_2^*) = \lambda_2^*.
 \end{equation}

We establish the existence and uniqueness of an equilibrium point
and provide equations characterizing it in Proposition
\ref{ExistenceOfEquilibriumPointDuopoly}, whose proof is deferred to
Appendix~D.

\begin{prop}
\label{ExistenceOfEquilibriumPointDuopoly} For any non-negative
price pair $(p_1, p_2)$, there exists a unique equilibrium point
$(\lambda_1^*, \lambda_2^*)$ of the user subscription dynamics in
the market with one incumbent. Moreover, $(\lambda_1^*,
\lambda_2^*)$ satisfies
\begin{equation} \label{eq:EquilibriumDuopoly}
\left\{
\begin{split}
\lambda_1^*=1-F\left(\frac{p_1}{q_1}\right) \text{, } \lambda_2^*=0,\;\;\;\;\;\;\;\;\;\;\;\;\;\;\;\;\;\;\;\; \text{ if }\frac{p_1}{q_1}\leq\frac{p_2}{g(0)},\\
\lambda_1^*=1-F\left(\theta_1^*\right) \text{, }
\lambda_2^*=F\left(\theta_1^*\right)-F\left(\theta_2^*\right),\;\;\;
\text{if }\frac{p_1}{q_1}>\frac{p_2}{g(0)},
\end{split}
\right.
\end{equation}
where $\theta_1^*=(p_1-p_2)/(q_1-g(\lambda_2^*))$ and
$\theta_2^*=p_2/g(\lambda_2^*)$.\hfill$\square$
\end{prop}

 Proposition
\ref{ExistenceOfEquilibriumPointDuopoly} indicates that, given any
prices $(p_1,p_2)$, the market shares of the two NSPs are uniquely
determined when the fraction of users subscribing to each NSP no
longer changes. Theoretically, this result ensures that if the NSPs
choose the optimal price (and also the entrant NSP selects the
optimal technology), then their corresponding profits will be
maximized, since the resulting outcome (e.g., equilibrium) is unique
and the NSPs face no uncertainty in the user subscription dynamics.
Proposition \ref{ExistenceOfEquilibriumPointDuopoly} also shows that
the structure of the equilibrium point depends on the relative
values of $p_1/q_1$ and $p_2/g(0)$. Specifically, if the price per
QoS of NSP $\mathcal{S}_1$ is always lower than or equal to that of
NSP $\mathcal{S}_2$, i.e., $p_1/q_1\leq p_2/g(0)$, then no users
subscribe to NSP $\mathcal{S}_2$ at the equilibrium point. On the
other hand, if NSP $\mathcal{S}_2$ offers a lower price per QoS to
its first subscriber than NSP $\mathcal{S}_1$ does, i.e.,
$p_1/q_1>p_2/g(0)$, then both NSP $\mathcal{S}_1$ and NSP
$\mathcal{S}_2$ may attract a positive fraction of subscribers. This
result regarding the price per QoS quantifies the
 necessary condition on prices that the entrant NSP
should set such that it can receive a positive revenue. We note
that, although it may be familiar to researchers and/or managers,
this result is important and relevant for the completeness of study,
as it rigorously characterizes the equilibrium outcome in the user
subscription dynamics and serves as the basis for both NSPs to make
pricing decisions and for the entrant to make technology decisions.
The importance of Proposition 3 can also be reflected in recent
works (e.g.,
\cite{JinSenGuerinHosanagarZhang}\cite{ManshaeiFreudigerFelegyhaziMarbachHubaux}\cite{ChauWangChiu}\cite{ShettyParekhWalrand})
which establish similar results under various settings.

We now investigate whether the user subscription dynamics specified
by (\ref{EvolutionDuopolyS1})--(\ref{EvolutionDuopolyS2_2})
stabilizes as time passes. As in the market with no incumbent, the
considered user subscription dynamics is guaranteed to converge to
the unique equilibrium when the QoS degradation of NSP
$\mathcal{S}_2$ is not too fast. In the following theorem, we
provide a sufficient condition for convergence.

\begin{theorem}
\label{ConvergenceOfEvolutionDuopoly} For any non-negative
price pair $(p_1,p_2)$, the user subscription dynamics specified by
(\ref{EvolutionDuopolyS1})--(\ref{EvolutionDuopolyS2_2}) converges to the unique equilibrium point
starting from any initial point $(\lambda_1^0,\lambda_2^0) \in \Lambda$ if
\begin{equation}
\label{ConvergenceConditionDuopoly}
\max_{\lambda_2\in[0,1]}\left\{-\frac{g'(\lambda_2)}{g(\lambda_2)}\cdot\frac{q_1}{q_1-g(\lambda_2)}\right\}<\frac{1}{K},
\end{equation}
where $K=
\max_{\alpha\in[0,\beta]}f(\alpha)\alpha.$
\end{theorem}
{\proof See Appendix~E.
\hfill$\square$}

%

Note that the condition (\ref{ConvergenceConditionDuopoly}) imposes
a more stringent requirement on the QoS function $g(\cdot)$ than the
condition (\ref{ConvergenceCondition}) does, since
$q_1/(q_1-g(\lambda_2)) > 1$ for all $\lambda_2 \in [0,1]$. However,
the condition (\ref{ConvergenceConditionDuopoly}) provides us with a
similar insight that, if QoS degradation is severe, the user
subscription dynamics may exhibit oscillation or divergence.


\subsection{Revenue Maximization}

We now study revenue maximization in the market with one incumbent.
In the economics literature, competition among a small number of
firms has been analyzed using game theory, following largely two
distinct approaches: Bertrand competition and Cournot competition
\cite{OsborneRubinstein}. In Bertrand competition, firms choose
prices independently while supplying quantities demanded at the
chosen prices. On the other hand, in Cournot competition, firms
choose quantities independently while prices are determined in the
markets to equate demand with the chosen quantities. In the case of
monopoly, whether the monopolist chooses the price or the quantity
does not affect the outcome since there is a one-to-one relationship
between the price and the quantity given a downward-sloping demand
function. This point was illustrated with our model in
Section~\ref{sec:RevenueMonopoly}. On the contrary, in the presence
of strategic interaction, whether firms choose prices or quantities
can affect the outcome significantly. For example, it is well-known
that identical firms producing a homogeneous good obtain zero profit
in the equilibrium of Bertrand competition while they obtain a
positive profit in the equilibrium of Cournot competition, if they
have a constant marginal cost of production and face a linear demand
function.

We first consider Bertrand competition between the two NSPs. Let
$\lambda_i^*(p_1,p_2)$ be the market share of NSP $\mathcal{S}_i$,
for $i=1,2$, at the unique equilibrium point of the considered user
subscription dynamics given a price pair $(p_1,p_2)$.
$\lambda_i^*(\cdot)$ can be interpreted as a demand function of NSP
$\mathcal{S}_i$, and the revenue of NSP $\mathcal{S}_i$ at the
equilibrium point  can be expressed as\footnote{Without causing
ambiguity, in the following analysis, we also express the revenue of
an NSP as a function of the fraction of subscribers.}
$R_i(p_1,p_2)=p_i\lambda_i^*(p_1,p_2)$, for $i=1,2$. Bertrand
competition in the market can be formulated as a non-cooperative
game specified by
\begin{equation}
\mathcal{G}_B=\left\{\mathcal{S}_i,R_i(p_1,p_2), p_i\in \mathbb{R}_+
\;|\;i=1,2\right\}.
 \end{equation}
A price pair $(p_1^*,p_2^*)$ is said to be a (pure) NE of $\mathcal{G}_B$
(or a Bertrand equilibrium) if it satisfies
\begin{equation}
R_i(p_i^*, p_{-i}^*)\geq R_i(p_i,
p_{-i}^*),\;\forall\,p_i\in\mathbb{R}_+,\forall \, i=1,2\,.
\end{equation}
It can be shown that, if a Bertrand equilibrium $(p_1^*,p_2^*)$ exists,
it must satisfy
\begin{equation}
0 < \frac{p_2^*}{g(0)} < \frac{p_1^*}{q_1} < \beta
\end{equation}
and $\lambda_i^*(p_1^*,p_2^*) \in (0,1)$ so that $R_i(p_1^*, p_2^*)
> 0$, for $i=1,2$. However, since the functions $\lambda_i^*(p_1,p_2)$, $i=1,2$, are defined
implicitly by \eqref{eq:EquilibriumDuopoly}, it is difficult to
provide a primitive condition on  $g(\cdot)$ that guarantees the
existence of a Bertrand equilibrium.

We now consider Cournot competition between the two NSPs. Let
$\lambda_i \in [0,1]$ be the market share chosen by NSP
$\mathcal{S}_i$, for $i=1,2$. Suppose that $\lambda_1 + \lambda_2
\leq 1$ so that the chosen market shares are feasible. Let
$p_i(\lambda_1, \lambda_2)$, $i=1,2$, be the prices that clear the
market, i.e., the prices that satisfy $\lambda_i =
\lambda_i^*(p_1(\lambda_1, \lambda_2),p_2(\lambda_1, \lambda_2))$
for $i=1,2$. Note first that, given a price pair $(p_1,p_2)$, if a
user $k$ subscribes to NSP $\mathcal{S}_1$, i.e., $\alpha_k q_1 -p_1
\geq \alpha_k g(\lambda_2) - p_2$ and $\alpha_k q_1  -p_1 \geq 0$,
then all the users whose valuations of QoS are larger than
$\alpha_k$ also subscribe to NSP $\mathcal{S}_1$. Also, if a user
$k$ subscribes to one of the NSPs, i.e., $\max \{\alpha_k q_1  -p_1,
\alpha_k g(\lambda_2) - p_2 \} \geq 0$, then all the users whose
valuations of QoS are larger than $\alpha_k$ also subscribe to one
of the NSPs. Therefore, realizing positive market shares $\lambda_1,
\lambda_2 > 0$ requires two types of marginal users whose valuations
of QoS are specified by $\alpha_{m,1}$ and $\alpha_{m,2}$ with
$\alpha_{m,1} > \alpha_{m,2}$. $\alpha_{m,1}$ is the valuation of
QoS of a marginal user that is indifferent between subscribing to
NSP $\mathcal{S}_1$ and NSP $\mathcal{S}_2$, while $\alpha_{m,2}$ is
the valuation of QoS of a marginal user that is indifferent between
subscribing to NSP $\mathcal{S}_2$ and neither. The expressions for
$\alpha_{m,1}$ and $\alpha_{m,2}$ that realize $(\lambda_1,
\lambda_2)$ such that $\lambda_1, \lambda_2 > 0$ and $\lambda_1 +
\lambda_2 \leq 1$ are given by
\begin{align}
 \label{AlphaFunctionAsFraction_1}
\alpha_{m,1}(\lambda_1, \lambda_2)&=z_{1}(\lambda_1)\triangleq F^{-1}(1-\lambda_1),\\
 \label{AlphaFunctionAsFraction_2}
\alpha_{m,2}(\lambda_1, \lambda_2)&=z_{2}(\lambda_1,\lambda_2)\triangleq F^{-1}(1-\lambda_1-\lambda_2).
 \end{align}
Also, by solving the indifference conditions, $\alpha_{m,1}q_1-p_1=\alpha_{m,1}g(\lambda_2)-p_2$
and $\alpha_{m,2}g(\lambda_2)-p_2=0$, we obtain a unique price pair that realizes
$(\lambda_1, \lambda_2)$ such that $\lambda_1, \lambda_2 > 0$ and
$\lambda_1 + \lambda_2 \leq 1$,
\begin{align}
 \nonumber
p_1(\lambda_1,\lambda_2)=&F^{-1}(1-\lambda_1)\left[q_1-g(\lambda_2)\right]\\
\label{PriceFunctionAsFraction_1}
&+F^{-1}(1-\lambda_1-\lambda_2)g(\lambda_2),\\
 \label{PriceFunctionAsFraction_2}
p_2(\lambda_1,\lambda_2)=&F^{-1}(1-\lambda_1-\lambda_2)g(\lambda_2).
 \end{align}
Note that the expressions
\eqref{AlphaFunctionAsFraction_1}--\eqref{PriceFunctionAsFraction_2}
are still valid even when $\lambda_i = 0$ for some $i=1,2$, although
uniqueness is no longer obtained. Hence, we can interpret
$p_i(\cdot)$, $i=1,2$, as a function defined on $\Lambda =
\{(\lambda_1,\lambda_2)\in\mathbb{R}_+^2\;|\;\lambda_1+\lambda_2\leq1\}$,
i.e., an inverse demand function in economics terminology. Then the
revenue of $\mathcal{S}_i$ when the NSPs choose $(\lambda_1,
\lambda_2) \in \Lambda$ is given by $R_i(\lambda_1,\lambda_2) =
\lambda_i p_i(\lambda_1,\lambda_2)$, for $i=1,2$. We define
$R_i(\lambda_1,\lambda_2)=0$, $i=1,2$, if $\lambda_1+\lambda_2>1$,
i.e., if the market shares chosen by the NSPs are infeasible.
Cournot competition in the market can be formulated as a
non-cooperative game specified by
\begin{equation}
\mathcal{G}_C=\left\{\mathcal{S}_i,R_i(\lambda_1,\lambda_2),
\lambda_i\in [0,1] \;|\;i=1,2\right\}.
 \end{equation}
A market share pair $(\lambda_1^{**},\lambda_2^{**})$ is said to be a (pure) NE of $\mathcal{G}_C$
(or a Cournot equilibrium) if it satisfies
\begin{equation}
R_i(\lambda_i^{**}, \lambda_{-i}^{**})\geq
R_i(\lambda_i, \lambda_{-i}^{**}),\;\forall\;\lambda_i\in
[0,1],\forall \; i=1,2\,.
\end{equation}
Note that $(1,1)$ is a NE of $\mathcal{G}_C$, which yields zero
profit to both NSPs. To eliminate this inefficient and
counterintuitive equilibrium, we restrict the strategy space of each
NSP to $[0,1)$. Deleting 1 from the strategy space can also be
justified by noting that $\lambda_i = 1$ is a weakly dominated
strategy for NSP $\mathcal{S}_i$, for $i=1,2$, since
$R_i(1,\lambda_{-i})
= 0 \leq R_i(\lambda_{i},\lambda_{-i})$ for all $(\lambda_{i},\lambda_{-i}) \in [0,1]^2$.%
\footnote{$\lambda_i \in [0,1]$ is a weakly (strictly) dominated
strategy for NSP $\mathcal{S}_i$ in $\mathcal{G}_C$
($\tilde{\mathcal{G}}_C$) if there exists another strategy
$\lambda'_i \in [0,1]$ such that $R_i(\lambda_{i},\lambda_{-i}) \leq
R_i(\lambda'_{i},\lambda_{-i})$ ($R_i(\lambda_{i},\lambda_{-i}) <
R_i(\lambda'_{i},\lambda_{-i})$) for all $\lambda_{-i} \in [0,1]$.}
We use $\tilde{\mathcal{G}}_C$ to represent the Cournot competition
game with the restricted strategy space $[0,1)$. The following lemma
bounds the market shares that solve the revenue maximization problem
of each NSP, when the PDF of the users' valuations of QoS satisfies
the non-increasing property as in Proposition
\ref{UpperLowerBoundsMonopoly}.

\begin{lem}
\label{SumMarketShareLessThanOne} Suppose that $f(\cdot)$ is
non-increasing on $[0,\beta]$. Let $\tilde{\lambda}_i(\lambda_{-i})$
be a market share that maximizes the revenue of NSP $\mathcal{S}_i$
provided that NSP $\mathcal{S}_{-i}$ chooses $\lambda_{-i} \in
[0,1)$, i.e., $\tilde{\lambda}_i(\lambda_{-i}) \in \arg
\max_{\lambda_i \in [0,1)}R_i(\lambda_i,\lambda_{-i})$. Then
$\tilde{\lambda}_i(\lambda_{-i}) \in (0,1/2]$ for all $\lambda_{-i}
\in [0,1)$, for all $i=1,2$. Moreover,
$\tilde{\lambda}_i(\lambda_{-i}) \neq 1/2$ if $\lambda_{-i} > 0$,
for $i=1,2$.
\end{lem}
{\proof The proof is similar to that of Proposition
\ref{UpperLowerBoundsMonopoly} and omitted for brevity.
\hfill$\square$}

Lemma \ref{SumMarketShareLessThanOne} implies that, when the
strategy space is specified as $[0,1)$ and $f(\cdot)$ satisfies the
non-increasing property, strategies $\lambda_i \in \{0\} \cup
(1/2,1)$ is strictly dominated for $i=1,2$. Hence, if a NE
$(\lambda_1^{**},\lambda_2^{**})$ of $\tilde{\mathcal{G}}_C$ exists,
then it must satisfy $(\lambda_1^{**}, \lambda_2^{**}) \in
(0,1/2)^2$, which yields positive revenues for both NSPs.
Furthermore, since a revenue-maximizing NSP never uses a strictly
dominated strategy, the set of NE of $\tilde{\mathcal{G}}_C$ is not
affected by restricting the strategy space to $[0,1/2]$. Based on
the discussion so far, we can provide a sufficient condition on
$f(\cdot)$ and $g(\cdot)$ that guarantees the existence of a NE of
$\tilde{\mathcal{G}}_C$.

\begin{theorem}
\label{ExistenceOfNashEquilibriumDuopoly} Suppose that $f(\cdot)$ is
non-increasing and continuously differentiable on
$[0,\beta]$.\footnote{We define the derivative of $f(\cdot)$ at 0
and $\beta$ using a one-sided limit as in footnote 3.} If $f(\cdot)$
and $g(\cdot)$ satisfy (\ref{SupermodularCondition_1}) and
(\ref{SupermodularCondition_2}) (shown on the top of the next page),
\begin{figure*}
\begin{align}
\nonumber
\left\{\frac{1}{f(F^{-1}(1-\lambda_1-\lambda_2))} + \frac{\lambda_1 f'(F^{-1}(1-\lambda_1-\lambda_2))}%
{[f(F^{-1}(1-\lambda_1-\lambda_2))]^3} \right\} g(\lambda_2) \qquad \\
\label{SupermodularCondition_1} + \left\{ F^{-1}(1-\lambda_1) -
\frac{\lambda_1}{f(F^{-1}(1-\lambda_1))} -
F^{-1}(1-\lambda_1-\lambda_2) +
\frac{\lambda_1}{f(F^{-1}(1-\lambda_1-\lambda_2))} \right\}
g'(\lambda_2) \geq 0
\end{align}
\begin{align}
\label{SupermodularCondition_2}
\left\{\frac{1}{f(F^{-1}(1-\lambda_1-\lambda_2))} + \frac{\lambda_2 f'(F^{-1}(1-\lambda_1-\lambda_2))}%
{[f(F^{-1}(1-\lambda_1-\lambda_2))]^3} \right\} g(\lambda_2)
 +\frac{\lambda_2}{f(F^{-1}(1-\lambda_1-\lambda_2))} g'(\lambda_2)
\geq 0
\end{align}
\hrulefill
\end{figure*}
for all $(\lambda_1,\lambda_2) \in[0,1/2]^2$, then the game
$\tilde{\mathcal{G}}_C$ has at least one NE.
\end{theorem}
{\proof See Appendix~F.
\hfill$\square$}

{We briefly discuss the conditions
(\ref{SupermodularCondition_1}) and (\ref{SupermodularCondition_2})
in Theorem \ref{ExistenceOfNashEquilibriumDuopoly} as follows. Under
these conditions, one NSP lowers its market share to maximize its
revenue when the other NSP increases its market share.
 In other words, if we treat $-\lambda_1$ as the
action of NSP $\mathcal{S}_1$, then the game becomes a supermodular
game and exhibits a strategic complementarity, i.e., the  NSPs'
strategies are compliments to each other \cite{Topkis}. Due to the
general distribution of the users' valuations of QoS, it is
difficult to characterize QoS functions satisfying the conditions
 (\ref{SupermodularCondition_1}) and
(\ref{SupermodularCondition_2}). Nevertheless, if we focus on the
uniform distribution of the users' valuations of QoS, the conditions
(\ref{SupermodularCondition_1}) and (\ref{SupermodularCondition_2})
coincide and reduce to $g(\lambda_2) + \lambda_2 g'(\lambda_2) \geq
0$, and thus  we obtain the following corollary.

\begin{corollary}
\label{ExistenceOfNashEquilibriumDuopolyLinear} Suppose that the
users' valuations of QoS are uniformly distributed, i.e.,
$f(\alpha)=1/\beta$ for $\alpha\in[0,\beta]$. If $g(\lambda_2) +
\lambda_2 g'(\lambda_2) \geq 0$ for all $\lambda_2\in[0,1/2]$, then
the game $\tilde{\mathcal{G}}_C$ has at least one NE.\hfill$\square$
\end{corollary}

Corollary \ref{ExistenceOfNashEquilibriumDuopolyLinear} states that
if the elasticity of the QoS provided by NSP $\mathcal{S}_2$ with
respect to the fraction of its subscribers is no larger than 1
(i.e., $- [g'(\lambda_2)\lambda_2/g(\lambda_2)] \leq 1$), the
Cournot competition game with the strategy space $[0,1)$ has at
least one NE. Note that the condition \eqref{ConvergenceCondition}
in Theorem \ref{ConvergenceOfEvolution} can be rewritten as
$g(\lambda_2) + K g'(\lambda_2)
> 0$ for all $\lambda_2 \in [0,1]$, where $K=\max_{\alpha\in[0,\beta]}f(\alpha)\alpha$.
Thus, the condition in Corollary
\ref{ExistenceOfNashEquilibriumDuopolyLinear} is analogous to the
sufficient conditions for convergence in that it requires that the
QoS provided by NSP $\mathcal{S}_2$ cannot degrade too fast with
respect to the fraction of subscribers. We
explain this point by considering a hypothetical scenario as
follows.  If  NSP $\mathcal{S}_1$ increases its action $-\lambda_1$
(i.e., lowers its market share) and the QoS provided by NSP
$\mathcal{S}_2$ degrades very rapidly when more users subscribe,
then  NSP $\mathcal{S}_2$ does not necessarily want to increase its
market share to maximize its revenue. This is because if NSP
$\mathcal{S}_2$ increases its market share, then its QoS may be very
low due to the severe degradation. Correspondingly, NSP
$\mathcal{S}_2$ has to charge a very low price to maintain the
increased market share and hence, its revenue may not be maximized.
Thus, strategic complementarity does not necessarily hold and an NE
may not necessarily exist. With a linearly-degrading QoS function
$g(\lambda_2)=\bar{q}_2-c\lambda_2$, we can obtain explicit
expressions of the NSPs' best responses as follows:
\begin{align}
\label{BestResponseS1}
&\mathcal{B}_1(\lambda_2)=\frac{q_1-\lambda_2(\bar{q}_2-c\lambda_2)}{2q_1},
\\
\label{BestResponseS2}
&\mathcal{B}_2(\lambda_1)=\frac{c(1-\lambda_1)+\bar{q}_2-\sqrt{\bar{q}_2^2
+ c^2(1-\lambda_1)^2 - c\bar{q}_2(1-\lambda_1)}}{3c}.
\end{align}
Moreover, we have the following corollary regarding the NE of the
game $\tilde{\mathcal{G}}_C$.

\begin{corollary}
\label{ExistenceOfNashEquilibriumDuopolyLinearQoS} If the users'
valuations of QoS are uniformly distributed, i.e.,
$f(\alpha)=1/\beta$ for $\alpha\in[0,\beta]$, and the QoS function
$g(\cdot)$ is linearly-degrading, i.e., $g(\lambda_2)=\bar{q}_2-c
\lambda_2$ for $\lambda_2\in[0,1]$, then the game
$\tilde{\mathcal{G}}_C$ has a unique NE, which can be reached
through the best response dynamics specified
in \eqref{BestResponseS1} and \eqref{BestResponseS2}.\\
\indent\textit{Proof:} By plugging $g(\lambda_2)=\bar{q}_2-c
\lambda_2$ into the condition $g(\lambda_2) + \lambda_2
g'(\lambda_2) \geq 0$, for all $\lambda_2\in[0,1/2]$, in Corollary
\ref{ExistenceOfNashEquilibriumDuopolyLinear}, we see that
$\bar{q}_2-2c\lambda_2\geq0$, for all $\lambda_2\in[0,1/2]$, since
$c<\bar{q}_2$. Hence, the existence of NE is proved. The uniqueness
of NE can be proved by solving the NE condition and checking that
only one point $(\lambda_1^{**},\lambda_2^{**})$ satisfies
$\lambda_1^{**}\in[0,1/2]$ and $\lambda_2^{**}\in[0,1/2]$. Since the
condition in Corollary \ref{ExistenceOfNashEquilibriumDuopolyLinear}
is satisfied, if we treat $-\lambda_1$ as the action of NSP
$\mathcal{S}_1$, then the game is a supermodular game with a unique
NE, to which the best response dynamics always converges
\cite{Topkis}. Thus, in the game $\tilde{\mathcal{G}}_C$, the best
response dynamics also converges to the NE. The details are omitted
for brevity. \hfill $\square$
\end{corollary}

If the NSPs do not have complete information regarding the market
(e.g.,  an NSP does not know how its competitor responds to its
price and market share change in the future), then an NE may not
necessarily be achieved directly and thus, we briefly discuss an
iterative process to reach a NE of the Cournot competition game.
Theorem \ref{ExistenceOfNashEquilibriumDuopoly} is based on the fact
that the Cournot competition game with the strategy space $[0,1/2]$
can be transformed to a supermodular game \cite{Topkis} when
(\ref{SupermodularCondition_1}) and (\ref{SupermodularCondition_2})
are satisfied. It is known that the largest and the smallest NE of a
supermodular game can be obtained by iterated strict dominance,
which uses the best response. A detailed analysis of this process
requires an explicit expression of the best response correspondence
of each NSP, which is not readily available without specific
assumptions on $f(\cdot)$ and $g(\cdot)$. If the users' valuations
of QoS are uniformly distributed and the QoS function $g(\cdot)$ is
linearly-degrading, then the NSPs $\mathcal{S}_1$ and
$\mathcal{S}_2$ can adopt the best responses given in
\eqref{BestResponseS1} and \eqref{BestResponseS2}, respectively,
until the unique NE is reached. As illustrated in Fig. \ref{timing},
there are three levels of time horizons. In the short-term horizon,
users make subscription decisions, whereas in the medium-term
horizon, the NSPs adjust their market shares based on the best
responses specified in \eqref{BestResponseS1} and
\eqref{BestResponseS2}. The long-term horizon is the life-span of
technologies. These different time horizons reflect that the NSPs do
not change their prices (determined by their desired market shares)
as often as the users change their subscription decisions, while the
NSPs change their prices more frequently than they make entry and
technology selection decisions. We assume that the medium-term
horizon is sufficiently longer than the short-term horizon such that
once the NSPs choose their prices (or desired market shares), the
equilibrium market shares are quickly reached by the users. At the
same time, we assume that the long-term horizon is sufficiently
longer than the medium-term horizon such that that the NSPs have
enough time to reach the NE of the game given their technologies. In
this sense, the best response dynamics is a reasonable approach to
reach the NE,  when the NSPs do not have sufficient information to
compute NE and thus cannot play it directly. Moreover, for
multi-stage decision making (i.e., leader-follower model) considered
in our study, it is common that decision makers adopt best response
dynamics to reach an equilibrium given the decisions made by their
``leaders''. For instance, a two-stage decision making process was
studied in \cite{ShettySchwartzWalrand}, where the authors neglected
the user subscription dynamics and derived the best-response prices
for Internet service providers.


\section{Entry and Technology Selection}

In the previous two sections, we have studied the user subscription
dynamics and revenue maximization for markets with no incumbents and
with one incumbent. In this section, we formalize the problem of
entry and technology selection as follows. Denote the set of
available options by
$\{\mathcal{T}_0,\mathcal{T}_1,\mathcal{T}_2\cdots\mathcal{T}_L\}$,
where $\mathcal{T}_0=\text{``Not Enter''}$ represents that the
entrant chooses not to enter the market.

We assume that the entrant knows the (expected) life-span of
technologies, and $k_l\geq0$ is the average cost per period over the
life-span associated with the technology $\mathcal{T}_l$, for
$l=1,2\cdots L$, i.e., total cost divided by the number of periods
in the life-span. Typically, the life-span of technologies is
sufficiently long compared to a short-term period of user
subscriptions, and hence the maximum average revenue per period is
approximately equal to the maximum per-period revenue at the
equilibrium (i.e., the revenue obtained during the first few
periods, e.g., time required for the user subscription dynamics to
converge, can be ignored) \cite{ShettySchwartzWalrand}. For the convenience of analysis, $k_l$ is
assumed to be independent of the fraction of subscribers served by
the entrant, once the technology $\mathcal{T}_l$ is selected and
deployed. Thus, the long-term profit during each period is
$R_{2,{l}}^*-k_l$, where $R_{2,{l}}^*$ is the per-period revenue
obtained by solving \eqref{RevenueNSPMonopolyS2Equivalent} for the
market with no incumbent and the per-period revenue at the Nash
equilibrium for the market with one incumbent. The subscript $l$
stresses that the revenue is associated with the technology
${\mathcal{T}_l}$ selected by the entrant, for $l=1,2\cdots L$. Note
that if $\mathcal{T}_0=\text{``Not Enter''}$ is selected, then the
associated cost $k_0=0$ and the corresponding revenue is zero.
Mathematically, the entry and technology selection problem can be
stated as
\begin{equation}
\label{TechnologySelectionMonopoly}
\begin{split}
\max_{{\mathcal{T}_l}\in\{\mathcal{T}_0,\mathcal{T}_1,\mathcal{T}_2\cdots\mathcal{T}_L\}}
\left(R_{2,{l}}^*-k_l\right),
\end{split}
\end{equation}
which can be solved by enumerating all the available options
$\{\mathcal{T}_0,\mathcal{T}_1,\mathcal{T}_2\cdots\mathcal{T}_L\}$.
For the market with no incumbent, if the users' valuations of QoS
are uniformly distributed and the QoS is linearly-degrading, then
the optimal revenue can be expressed in a closed form and hence, the
entry and technology selection problem can be explicitly solved
based on \eqref{TechnologySelectionMonopoly}.

\begin{table}[!t]
\centering \caption{Network Parameters} \centering
\begin{tabular}{|l|l|}  \hline
 \textbf{Parameter} & \textbf{\;\;\;\;Value}\\
 \hline \multirow{3}{*}{Broadband factor} & Incumbent: 2
 \\ & Entrant (split): 1.90
 \\ & Entrant (common): 1.85
 \\ \hline
 Activity ratio & 0.8\\
 \hline {Macrocell capacity} & 0.5\\
 \hline
 {Degradation coefficient} & 0.4
 \\
 \hline
  Fraction outside  & 0.3 \\
 \hline
 \end{tabular}
 \label{SimulationEnvironment}
\end{table}

\begin{figure}[t]
\begin{center}
\includegraphics[width=7.7cm]{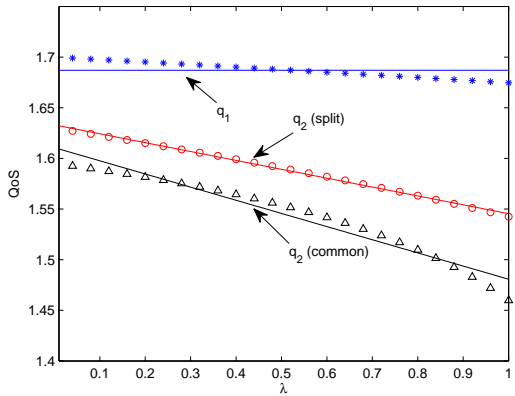}
\caption{QoS function (with a unit of ``bit/sec'') and
approximation. The actual and approximated QoS functions are plotted
in markers and solid lines, respectively. Approximated QoS
functions: $q_1=1.687$, $q_2=g(\lambda_2)=1.633-0.088\lambda_2$
(split), $q_2=g(\lambda_2)=1.611-0.129\lambda_2$ (common).}
\label{qos_approx}
\end{center}
\end{figure}

\section{Numerical Results}

In this section, we provide numerical results to complement the
analysis. For simplicity, we focus on two spectrum sharing schemes,
namely, ``split'' and `'common'', which are available for the
entrant. Mathematically, the set of available options for the
entrant can be denoted by $\{\mathcal{T}_0=``\text{Not
Enter''},\mathcal{T}_1=\text{``Split''},\mathcal{T}_2=\text{``Common''}\}$.
Since we mainly focus on the entry and technology selection for the
entrant, we assume as an example that the incumbent uses the
``split'' spectrum sharing scheme for its femtocells and macrocells.
Note that we can carry out a similar analysis while assuming that
the incumbent operates under the ``common'' spectrum sharing scheme,
although the specific results of entry and technology selection for
the entrant may be different. We also assume that the incumbent has
three times the bandwidth as the entrant, which reflects the fact
that the incumbent has more resources than the entrant, and that the
users' valuations of QoS are uniformly distributed in $[0,1]$, i.e.,
$f(\alpha)=1$ for $\alpha\in[0,1]$.

\begin{figure*}[!t]
  \centering
\subfigure[Revenue versus the market share with no incumbent. The
revenues based on the actual and approximated QoS functions are
plotted in circles and solid line,
respectively.]{\label{revise_mono_revenue_lambda_2}\includegraphics[width=5.9cm]{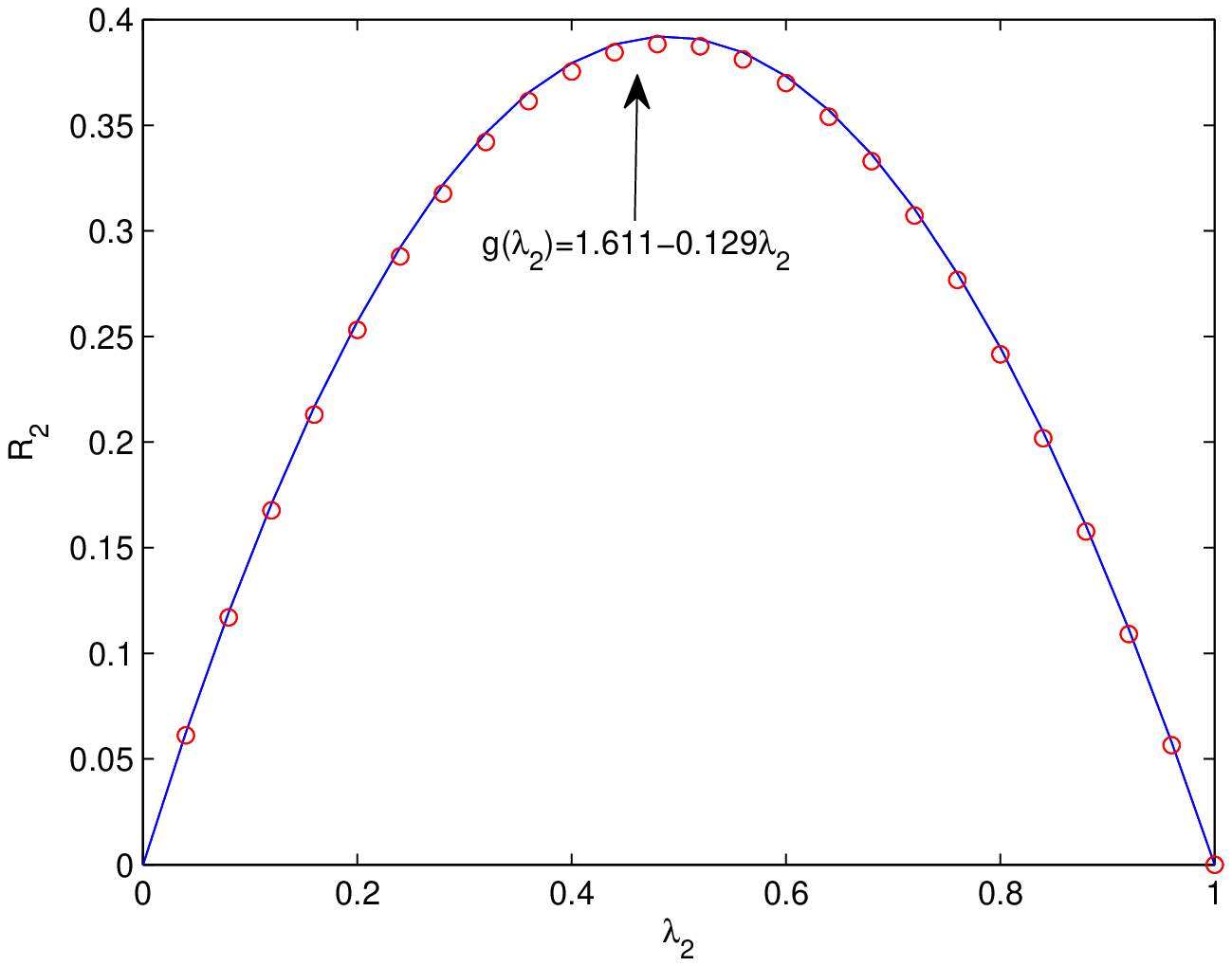}}
\subfigure[Convergence of the user subscription dynamics with no
incumbent. $p_2=1.2$ and $q_2=g(\lambda_2)=1.633-0.088\lambda_2$
(split).]{\label{revise_mono_convergence}\includegraphics[width=5.9cm]{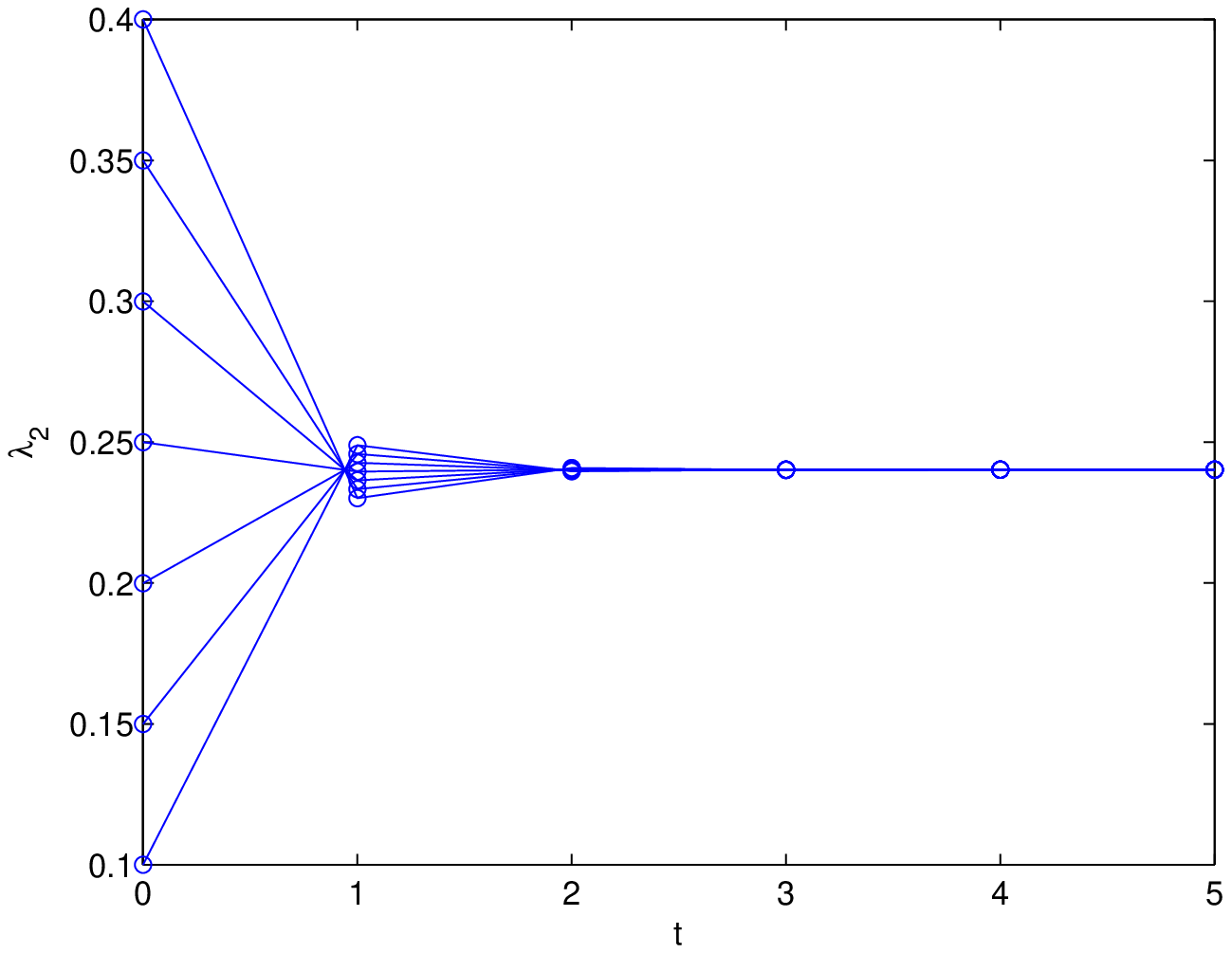}}
\subfigure[Revenue versus the market share with no incumbent.
Circle: common; triangle:
split.]{\label{new_revise_mono_revenue_lambda}\includegraphics[width=5.9cm]{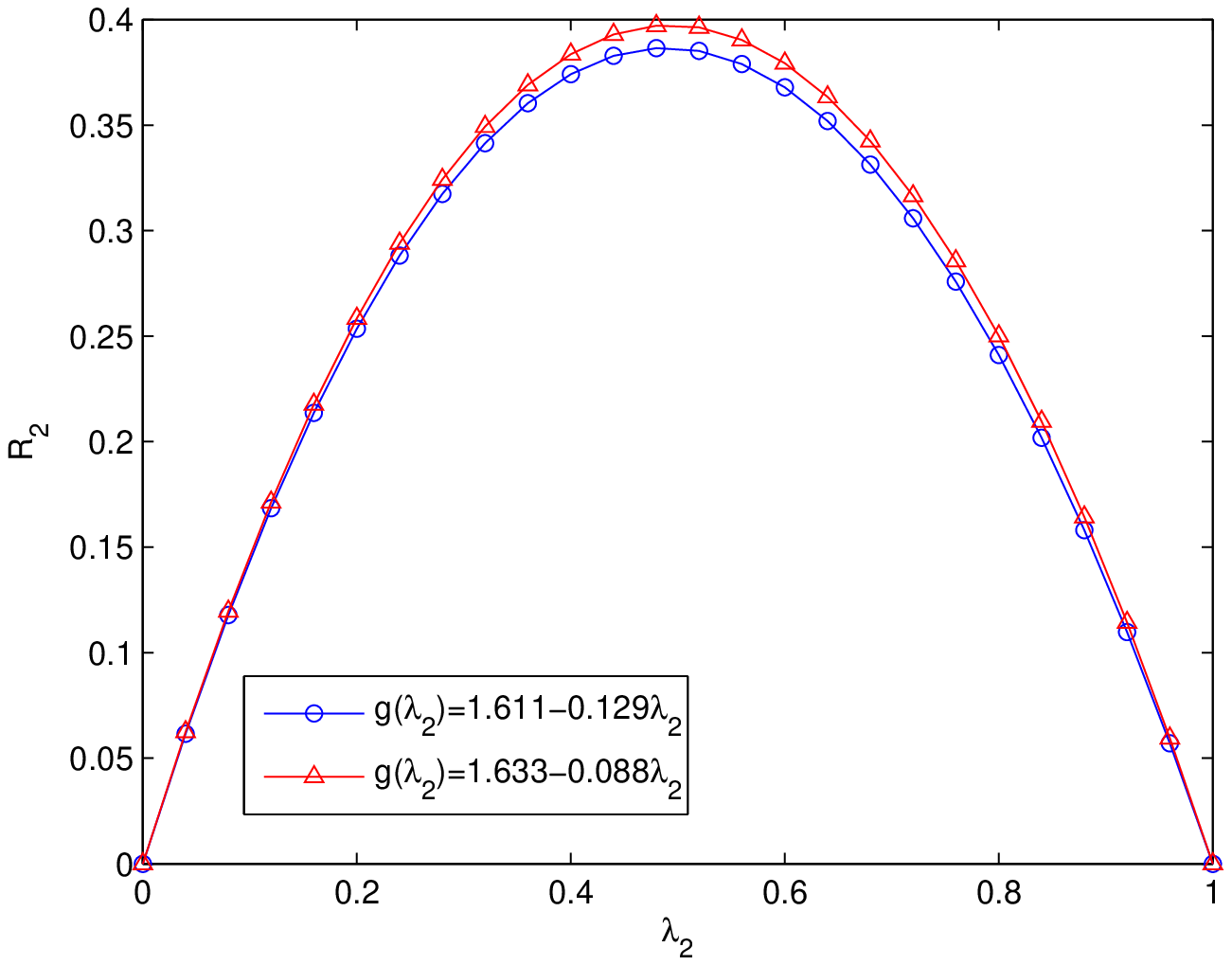}}
\caption{Market with no incumbent.}
\end{figure*}

Although our analysis applies to any QoS metric and QoS function
satisfying Assumption~1, we shall explicitly consider ``average
(normalized) throughput'', which has a unit of bit/sec and measures
the average transmission rate offered by the NSPs. In a femtocell
market, if a user subscribes to either of the two NSPs, it can use
femtocell at home and macrocell base stations while staying
outdoors. Hence, to derive the average throughput, both the users'
outdoor and indoor accesses need to be considered
\cite{ShettyParekhWalrand} and the average throughput can be
expressed as
\begin{equation}
\label{AverageThroughput}
\begin{split}
\mathbb{E}(T_f) = (1-f_o)\cdot\mathbb{E}(T_b) + f_o\cdot\mathbb{E}(T_m),
\end{split}
\end{equation}
where $f_o$ the fraction of time that users spend outdoors, $T_b$ is
the throughput obtained by a user from his broadband connection via
the femtocell, and $T_m$ is the throughput obtained via macrocells.
By considering the users' time-varying positions, transmitted data
sizes, and network congestions, the authors in
\cite{ShettyParekhWalrand} derived an explicit expression of
\eqref{AverageThroughput} which, due to its complexity, is not shown
here. In Table \ref{SimulationEnvironment}, we show the network
parameters, the meaning of which can be found in
\cite{ShettyParekhWalrand}. We compute the NSPs' QoS  functions
based on \eqref{AverageThroughput} derived in
\cite{ShettyParekhWalrand} and plot them in Fig. \ref{qos_approx}.
Using minimum mean square error fitting, we approximate the QoS
provided by NSP $\mathcal{S}_1$ using a constant and the QoS
provided by NSP $\mathcal{S}_2$ using an affine function (i.e.,
linearly-degrading QoS).\footnote{Note that our analysis does not
require the QoS function of NSP $\mathcal{S}_2$ to be
linearly-degrading that the affine approximation is applied mainly
because it allows us to derive more specific analytical results.}
The approximated QoS functions are shown as solid lines in Fig.
\ref{qos_approx}. We see from Fig. \ref{qos_approx} that although
the QoS function of NSP $\mathcal{S}_1$ is also decreasing in the
number of subscribers, its slope is much less than that of NSP
$\mathcal{S}_2$'s QoS function and approximating it using a constant
still stays close to the actual QoS (within $2\%$). It is also
observed from Fig. \ref{qos_approx} that the QoS provided by NSP
$\mathcal{S}_2$ satisfies the property of non-increasing in the
number of subscribers. While approximating the QoS using an affine
function when NSP $\mathcal{S}_2$ uses the ``split'' spectrum
sharing scheme is fairly accurate, the affine approximation is not
close to the actual QoS if the ``common'' spectrum sharing scheme is
used. Nevertheless, it can be seen from Fig.
\ref{revise_mono_revenue_lambda_2} that the revenue obtained for the
(approximated) linearly-degrading QoS function is very close
 to that obtained for the actual QoS
function (within $1\%$).\footnote{Note that, for the incumbent and
for the entrant using a split spectrum sharing scheme, the revenues
obtained based on approximated QoS functions are also very close to
those obtained based on the actual QoS functions, although they are
not shown in Fig. \ref{revise_mono_revenue_lambda_2}.} Then, by
using the marginal user principle, the optimal prices obtained based
on the approximated linearly-degrading and actual QoS functions are
$0.837$ and $0.839$, respectively, which are very close  to each
other (within $1\%$). Thus, approximating the QoS function using an
affine function is sufficiently accurate for the purpose of
maximizing the revenue, and our previous analysis based on
linearly-degrading QoS functions can be applied without losing much
accuracy.

\subsection{With no incumbent}

We first consider a market  with no incumbent. Fig.
\ref{revise_mono_convergence} illustrates the convergence of the
user subscription dynamics for a particular price $p_2=1.2$, when
the entrant uses the split spectrum sharing technology. Note that
given any price $p_2\geq0$, convergence will always be obtained,
since the QoS function satisfies the sufficient condition for
convergence given  in Theorem \ref{ConvergenceOfEvolution}. Note
that  convergence can also be observed if the entrant uses the
common spectrum sharing technology, which is not shown in the paper
for brevity. Fig. \ref{new_revise_mono_revenue_lambda} verifies
Proposition \ref{UpperLowerBoundsMonopoly} that the optimal market
share maximizing the revenue of NSP $\mathcal{S}_2$ is upper bounded
by $1/2$. We also observe from Fig.
\ref{new_revise_mono_revenue_lambda} that the split spectrum sharing
technology can yield a higher revenue for NSP $\mathcal{S}_2$, since
it provides a higher QoS, compared to the common spectrum sharing
technology. Nevertheless, to select the technology that maximizes
the entrant's long-term profit, it also needs to take into account
the cost associated with the employed technology (such as developing
and implementing spectrum sharing protocol stacks, build base
stations that complies with employed technology, etc.). We
illustrate in the upper plot of Fig. \ref{tech_selection} the
technology selection made by the entrant for different costs $k_1$
and $k_2$. It shows that, even though the split spectrum sharing
technology (``split'') offers a higher QoS than the common spectrum
sharing technology (``common'') at any number of subscribers, the
entrant may still select the ``common'' technology if the associated
cost is sufficiently lower than that associated with the ``split''
technology. This result quantifies the condition under which the
entrant should select ``split'' or ``common'', and serves as a
quantitative guidance for the entrant to choose a spectrum sharing
technology and maximize its profit.

\begin{figure}[!t]
\begin{center}
\includegraphics[width=7.7cm]{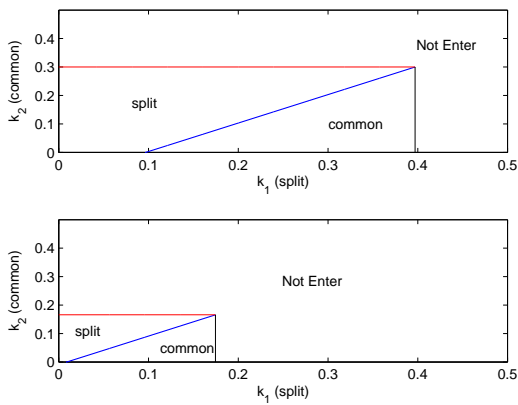}
\caption{Technology selection for different costs. Top: with no
incumbent; bottom: with one incumbent.} \label{tech_selection}
\end{center}
\end{figure}


\subsection{With one incumbent}

\begin{figure*}[!t]
  \centering
\subfigure[Convergence of market shares under the best-response
dynamics.]{\label{revise_duopoly_lambda}\includegraphics[width=5.9cm]{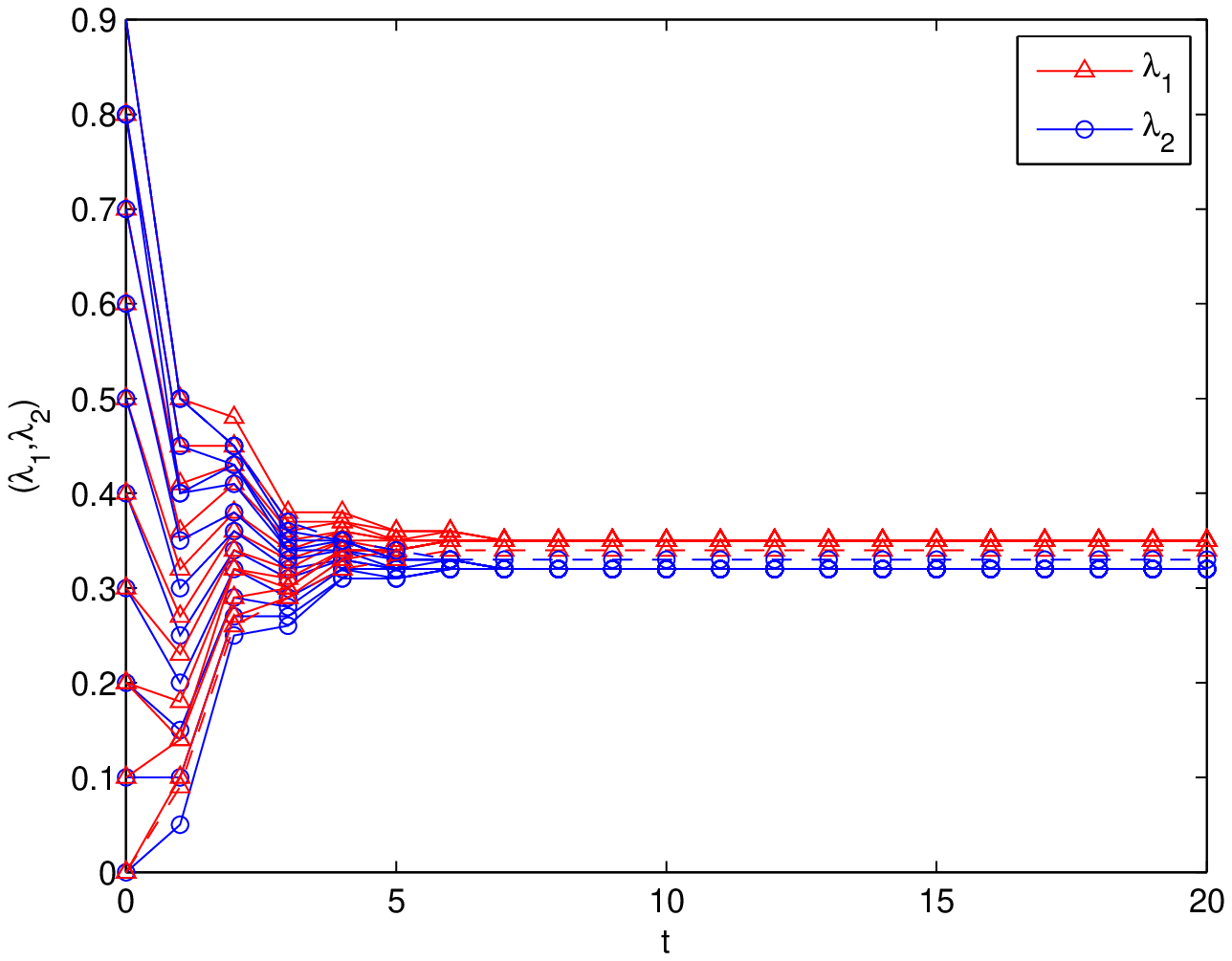}}
\subfigure[Convergence of prices under the best-response
dynamics.]{\label{revise_duopoly_price}\includegraphics[width=5.9cm]{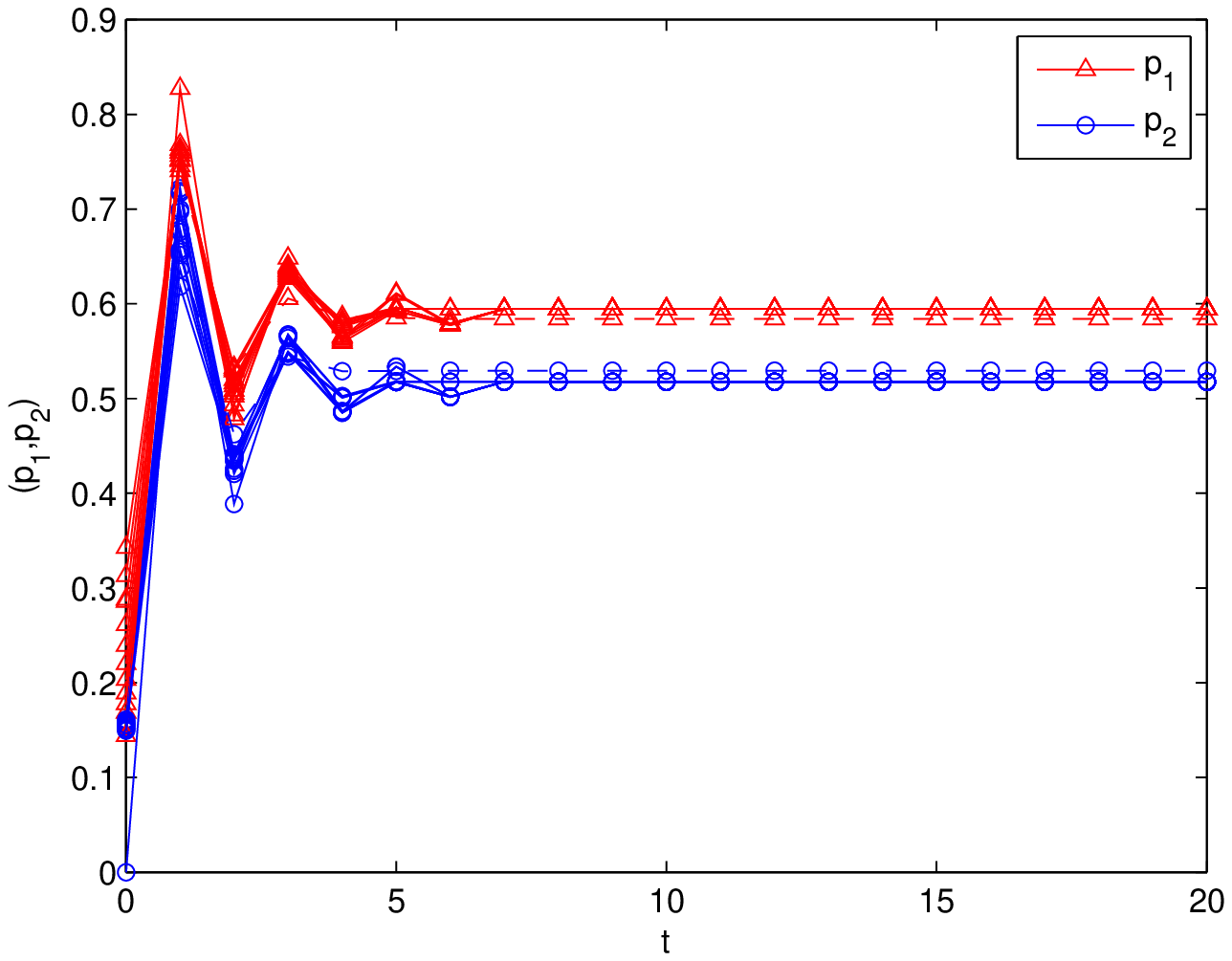}}
\subfigure[Convergence of revenues under the best-response dynamics.
]{\label{revise_duopoly_revenue}\includegraphics[width=5.9cm]{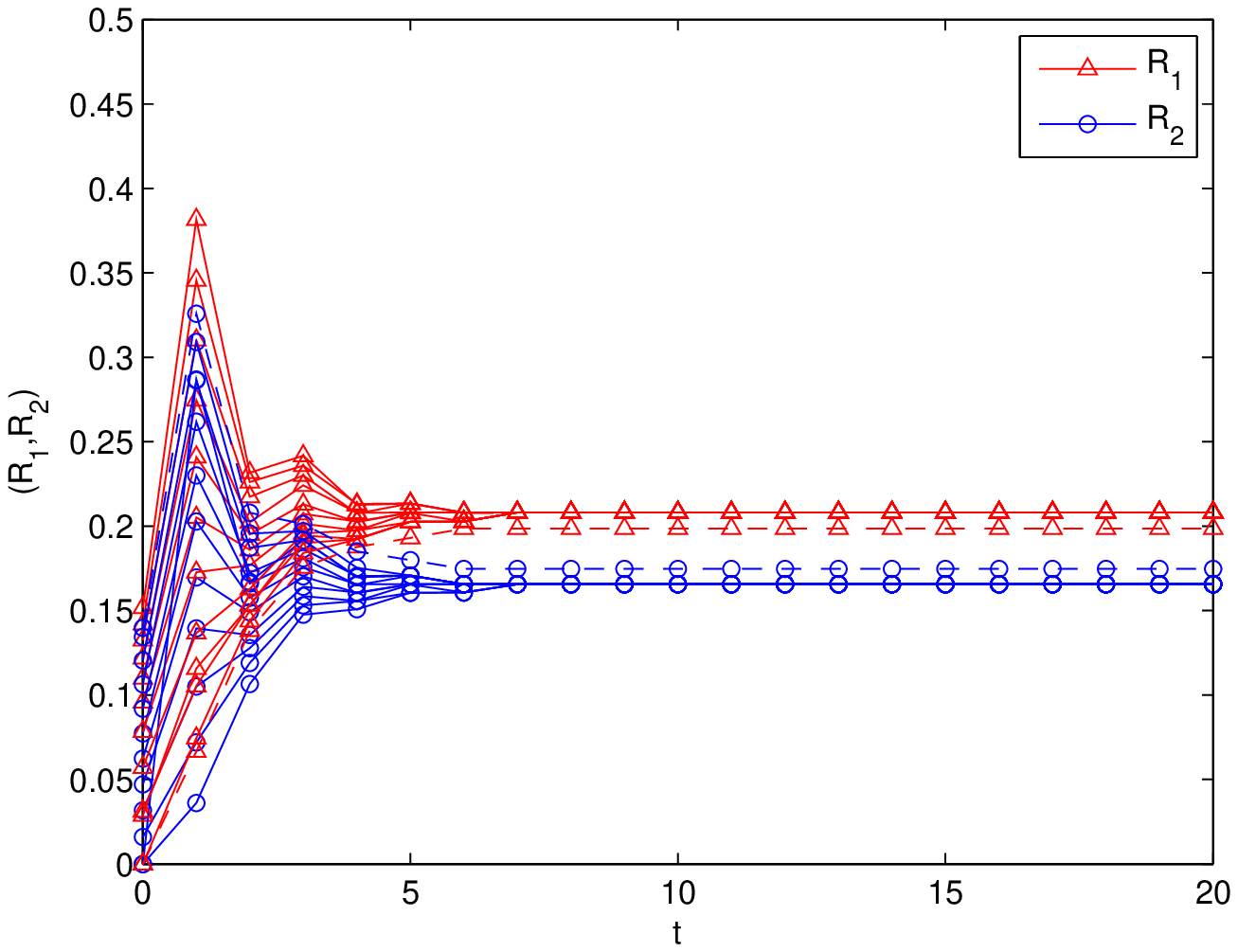}}
\caption{Market with one incumbent. Dashed and solid lines represent
that the entrant uses the split and common spectrum sharing
technologies, respectively.}
\end{figure*}

Now, we provide some numerical results regarding the market share
competition game. The QoS functions we use in the numerical results
are approximate affine functions, rather than the actual QoS
functions. Note, however, that we can obtain almost the same results
if we use the actual QoS functions, since the affine approximation
is sufficiently accurate for analyzing the revenue.
The convergence of the considered user subscription dynamics is
similar with that in the market with no incumbent, and hence it is
omitted due to the space limitations. Starting from different
initial points, Figs. \ref{revise_duopoly_lambda},
\ref{revise_duopoly_price}, and \ref{revise_duopoly_revenue}
 show the convergence of market shares, prices, and revenues, respectively, when
 both NSPs $\mathcal{S}_1$ and $\mathcal{S}_2$ update their
 market shares by choosing their best responses to the market share
 of the other NSP in the previous period.
Since the considered QoS functions satisfy the conditions in
Corollary \ref{ExistenceOfNashEquilibriumDuopolyLinearQoS}, the
Cournot competition game with the strategy space $[0,1/2]$ has a
unique NE, as verified in Fig. \ref{revise_duopoly_lambda}.
Moreover, Fig. \ref{revise_duopoly_lambda}  is consistent with Lemma
\ref{SumMarketShareLessThanOne} as the best-response market shares
of both NSPs do not exceed 1/2. It can also be observed from Figs.
\ref{revise_duopoly_lambda} and \ref{revise_duopoly_revenue}
 that if NSP $\mathcal{S}_2$ uses the common spectrum sharing technology
 that provides  a lower QoS
  (shown in solid lines) compared to that provided by the split spectrum
  sharing technology (shown in dashed lines), it obtains a smaller revenue, while NSP
$\mathcal{S}_1$ obtains a higher revenue. This is because when NSP
$\mathcal{S}_2$ has a lower QoS, it tries to maintain its market
share by lowering its price to compensate for the lower QoS, as can
be seen from Fig. \ref{revise_duopoly_price}. By comparing the
entrant's technology selection in a market with no and one incumbent
(shown in Fig. \ref{tech_selection}), we notice that competition
from the incumbent sets a barrier for the entrant to enter the
market. That is, the presence of an incumbent lowers the cost
threshold  under which the entrant earns positive profit from
entering the market. Moreover, our analysis provides the entrant
with a quantitative guideline as to whether to enter a
communications market and which technology to select such that it
can maximize its long-term profit.

\section{Conclusion}


Focusing on a femtocell market, we studied in this paper the problem
of long-term entry and spectrum sharing scheme decision for an
entrant. To address the long-term decision, we also studied two
related problems: the entrant's medium-term pricing decisions and
the users' short-term subscription decisions. We considered two
scenarios, one with no incumbent and the other with one incumbent.
In each scenario, we constructed the user subscription dynamics
based on static learning, and showed that there exists a unique
equilibrium point of the user subscription dynamics at which the
number of subscribers does not change. We provided a sufficient
condition on the entrant's QoS function that ensures the global
convergence of the user subscription dynamics. We also examined the
revenue maximization problem by the NSPs. With no incumbent in the
market, we derived upper and lower bounds on the optimal price and
the resulting market share that maximize the entrant' revenue, for a
non-increasing PDF of the users' valuations of QoS. With one
incumbent in the market, we studied competition between the two
NSPs, primarily focusing on market share competition. We modeled the
NSPs as strategic players in a non-cooperative game where each NSP
aims to maximize its own revenue by choosing its market share. We
obtained a sufficient condition that ensures the existence of at
least one NE of the game. Finally, we formalized the problem of
entry and spectrum sharing scheme selection for the entrant and
provided numerical results to complete our analysis. Our analysis
provides the entrant with a quantitative guideline as to whether to
enter a communications market and which technology to select such
that it can maximize its long-term profit. Future research
directions include, but are not limited to: (1) multiple incumbents
and/or entrants in the market; (2) general QoS functions for the
incumbents; and (3) social welfare maximization.

\appendices
\section{Proof of Proposition \ref{ExistenceOfEquilibriumPointMonopoly}}
To facilitate the proof, we first define
$\tilde{h}_m(\lambda_2)=h_m(\lambda_2)-\lambda_2$ for
$\lambda_2\in[0,1]$, where $h_m(\cdot)$ is defined in
(\ref{EvolutionProcess2}). By Definition 1, $\lambda_2^*$
is an equilibrium point if and only if it is a root of $\tilde{h}_m(\cdot)$.
Hence, it suffices to show that $\tilde{h}_m(\cdot)$ has a unique root
on its domain.

Suppose $p_2 = 0$. Then $h_m(\lambda_2) = 1$ for all $\lambda_2 \in [0,1]$.
Thus, $\tilde{h}_m(1) = 0$ while $\tilde{h}_m(\lambda_2) > 0$ for all $\lambda_2
\in [0,1)$. This implies that $\lambda_2 = 1$ is the unique root of $\tilde{h}_m(\cdot)$.

Suppose $0 < p_2 \leq \beta g(1)$. By the fundamental theorem of
calculus, $F(\cdot)$ is differentiable on $(0, \beta)$ with
$F'(\cdot) = f(\cdot)$. By applying the chain rule, we have
\begin{equation} \label{DerivativeHm2}
\frac{d{h}_m(\lambda_2)}{d\lambda_2} =
f\left(\frac{p_2}{g(\lambda_2)}\right)\frac{g'(\lambda_2)p_2}{[g(\lambda_2)]^2}
\end{equation}
for all $\lambda_2 \in (0,1)$. By Assumption 1, $g'(\cdot) \leq 0$ on $(0,1)$,
and thus $d{h}_m(\cdot)/d\lambda_2 \leq 0$ on $(0,1)$.
Since $d\tilde{h}_m(\cdot)/d\lambda_2 = d{h}_m(\cdot)/d\lambda_2 - 1$,
$\tilde{h}_m(\cdot)$ is strictly decreasing on $[0,1]$. Next,
we note that $\tilde{h}_m(0)=h_m(0)=1-F(p_2/g(0))\geq0$ and
$\tilde{h}_m(1)=-F(p_2/g(1))<0$.
Since $\tilde{h}_m(\cdot)$ is continuous on $[0,1]$, we obtain a unique root
of $\tilde{h}_m(\cdot)$ on $[0,1]$ by applying the intermediate value
theorem.

Suppose $\beta g(1) < p_2 < \beta g(0)$. Let $\bar{\lambda}_2 = \min \{ \lambda_2 \in [0,1] |
\beta g(\lambda_2) \leq p_2 \}$. Note that $\bar{\lambda}_2 \in (0,1)$.
Also, $h_m(\lambda_2) = 0$ for all $\lambda_2 \geq \bar{\lambda}_2$, and thus
$\tilde{h}_m(\lambda_2) = - \lambda_2 < 0$ for all $\lambda_2 \geq \bar{\lambda}_2$.
Hence, if a root of $\tilde{h}_m(\cdot)$ exists, it must be in $[0,\bar{\lambda}_2)$.
By applying a similar argument as above to the interval $[0,\bar{\lambda}_2]$,
we can show that $\tilde{h}_m(\cdot)$ has a unique root on $[0,\bar{\lambda}_2]$.

Suppose $p_2 \geq \beta g(0)$. Then $h_m(\lambda_2) = 0$ for all $\lambda_2 \in [0,1]$.
Thus, $\tilde{h}_m(0) = 0$ while $\tilde{h}_m(\lambda_2) < 0$ for all $\lambda_2
\in (0,1]$. This implies that $\lambda_2 = 0$ is the unique root of $\tilde{h}_m(\cdot)$.\hfill$\blacksquare$

\section{Proof of Theorem \ref{ConvergenceOfEvolution}}
We prove the convergence of the user subscription dynamics in the
market with no incumbent based on the contraction mapping theorem.

\textit{Definition 4 \cite{BertsekasTsitsiklis}:} A mapping
$\mathbf{T}:\,\mathcal{X}\to\mathcal{X}$, where $\mathcal{X}$ is a
closed subset of $\mathbb{R}^{n}$, is called a contraction if there
is a real number $\kappa\in[0,1)$ such that
\begin{equation}
\label{DefinitionContraction}
\|\mathbf{T}(x_1)-\mathbf{T}(x_2)\|\leq
\kappa\|x_1-x_2\|,\;\;\;\forall\;x_1,x_2\in\mathcal{X},
\end{equation}
where $\|\cdot\|$ is some norm defined on $\mathcal{X}$.


Proposition 1.1 in Chapter 3 of \cite{BertsekasTsitsiklis} shows an
important property of a contraction mapping $\mathbf{T}$ that the
update sequence generated by $x^{t+1}=\mathbf{T}(x^t)$, $t
=0,1,2,\ldots$, converges to a fixed point $x^*$ satisfying
$\mathbf{T}(x^*)=x^*$ starting from any initial value $x^0 \in
\mathcal{X}$. To prove Theorem \ref{ConvergenceOfEvolution}, we
shall show that the function $h_m(\cdot)$, defined in
(\ref{EvolutionProcess2}), is a contraction mapping on $[0,1]$ with
respect to the absolute value norm if the condition
(\ref{ConvergenceCondition}) is satisfied.

Suppose $p_2 = 0$. Then $h_m(\lambda_2) = 1$ for all $\lambda_2 \in [0,1]$,
and thus $h_m(\cdot)$ is a contraction with $\kappa = 0$.

Suppose $p_2 > 0$.
Let $\lambda_{2,a}$ and $\lambda_{2,b}$ be two different real
numbers arbitrarily chosen from the interval $[0,1]$, and suppose
without loss of generality that $\lambda_{2,a}>\lambda_{2,b}$.
We will show that
\begin{eqnarray}
\label{InequalityAppendixB1}
|h_m(\lambda_{2,a})-h_m(\lambda_{2,b})| \leq \kappa_m |\lambda_{2,a}-\lambda_{2,b}|,
\end{eqnarray}
where $\kappa_m=K \cdot \max_{\lambda_2\in[0,1]} \{-g'(\lambda_{2})/g(\lambda_{2}) \}$.
Then the condition (\ref{ConvergenceCondition}) implies that $\kappa_m \in [0,1)$,
establishing that $h_m(\cdot)$ is a contraction.
Since $0< p_2/g(\lambda_{2,b}) \leq p_2/g(\lambda_{2,a})$, we can
consider three cases.

\textit{Case 1 ($p_2/g(\lambda_{2,a}),  p_2/g(\lambda_{2,b}) < \beta$):}
Note that $h_m$ is continuous on $[0,1]$ and differentiable on
$(\lambda_{2,b},\lambda_{2,a})$. Hence, by the mean value theorem, there
exists $\lambda_{2,c}\in(\lambda_{2,b},\lambda_{2,a})$ such that
\begin{equation}
\label{InequalityAppendixB}
h'_m(\lambda_{2,c}) = \frac{h_m(\lambda_{2,a})-h_m(\lambda_{2,b})}{\lambda_{2,a}-\lambda_{2,b}}.
\end{equation}
Then we obtain
\begin{eqnarray}
\label{AppendixBLong2}
& &|h_m(\lambda_{2,a})-h_m(\lambda_{2,b})|\\
&=&\left|f\left(\frac{p_2}{g(\lambda_{2,c})}\right)\frac{p_2}
{g(\lambda_{2,c})}\frac{g'(\lambda_{2,c})}{g(\lambda_{2,c})}\right| |\lambda_{2,a}-\lambda_{2,b}|\\
\label{AppendixBLongAdditional_1} &\leq& \kappa_m
|\lambda_{2,a}-\lambda_{2,b}|.
\end{eqnarray}

\textit{Case 2 ($p_2/g(\lambda_{2,b}) < \beta \leq p_2/g(\lambda_{2,a})$):}
Let $\bar{\lambda}_2 = \min \{ \lambda_2 \in [0,1] | \ \beta g(\lambda_2) \leq p_2 \}$.
Note that $\lambda_{2,b} < \bar{\lambda}_2 \leq \lambda_{2,a}$. Applying
the mean value theorem to $h_m(\cdot)$ on the interval $[\lambda_{2,b}, \bar{\lambda}_2]$ yields
\begin{equation}
\label{InequalityAppendixB2}
|h_m(\bar{\lambda}_2)-h_m(\lambda_{2,b})| \leq \kappa_m |\bar{\lambda}_2-\lambda_{2,b}|.
\end{equation}
Since $h_m(\bar{\lambda}_2) = h_m(\lambda_{2,a}) = 0$ and $\kappa_m \geq 0$,
we obtain
\begin{equation}
\begin{split}
\label{InequalityAppendixB3} |h_m(\lambda_{2,a})-h_m(\lambda_{2,b})|
&= |h_m(\bar{\lambda}_2)-h_m(\lambda_{2,b})|\\ &\leq \kappa_m
|\bar{\lambda}_2-\lambda_{2,b}|\\
 &\leq \kappa_m
|\lambda_{2,a}-\lambda_{2,b}|.
\end{split}
\end{equation}

\textit{Case 3 ($p_2/g(\lambda_{2,a}),  p_2/g(\lambda_{2,b}) \geq
\beta$):} In this case, $h_m(\lambda_{2,a})=h_m(\lambda_{2,b})=0$,
and \eqref{InequalityAppendixB1} is trivially
satisfied.\hfill$\blacksquare$

\section{Proof of Proposition \ref{UpperLowerBoundsMonopoly}}

To prove Proposition \ref{UpperLowerBoundsMonopoly}, we shall show
that a solution $\lambda_2^{**}$ of $\max_{\lambda_2\in[0,1]}
F^{-1}(1-\lambda_2) g(\lambda_2) \lambda_2$ satisfies $0<
\lambda_2^{**} \leq 1/2$. Then $F^{-1}(1/2)\leq\alpha^*<\beta$
follows from the relationship $\alpha^* = F^{-1}(1-\lambda_2^{**})$
and $F^{-1}(1/2) g(1/2)\leq p_2^*<\beta g(0)$ from $p_2^* = \alpha^*
g(1-F(\alpha^*))$. Note first that $\lambda_2^{**}$ cannot be zero
because the maximum revenue is positive. Thus, it remains to show
$\lambda_2^{**} \leq 1/2$.

Let $z(\lambda_2)=F^{-1}(1-\lambda_2)$ for $\lambda_2 \in [0,1]$.
Then $z(\cdot)$ is 
differentiable on $(0,1)$ with the derivative $z'(\lambda_2) = dz(\lambda_2) / d\lambda_2
=-1/f(z(\lambda_2))$. Note that $z'(\lambda_2) < 0$ for all $\lambda_2 \in (0,1)$,
which implies that $z(\cdot)$ is strictly decreasing on $[0,1]$.
Also, since $f(\cdot)$ is non-increasing on $[0,\beta]$,
$z'(\cdot)$ is non-decreasing on $(0,1)$. The revenue of NSP $\mathcal{S}_2$
can be expressed as a function of $\lambda_2$, $R_2(\lambda_2)=\lambda_2g(\lambda_2)z(\lambda_2)$.
Since $R_2(\cdot)$ is differentiable and $\lambda_2^{**} \in (0,1)$
($\lambda_2 = 1$ yields zero revenue and thus cannot be optimal), the first-order necessary condition implies that
\begin{equation}
\begin{split}
\label{DerivativeRevenueLamda2NSPMonopolyS2}
\frac{d R_2(\lambda_2^{**})}{d\lambda_2}=&\left[z(\lambda_2^{**})+\lambda_2^{**}
z'(\lambda_2^{**})\right]g(\lambda_2^{**})\\&+\lambda_2^{**}z(\lambda_2^{**})g'(\lambda_2^{**})
= 0.
\end{split}
\end{equation}
Note that $\lambda_2^{**}z(\lambda_2^{**})g'(\lambda_2^{**}) \leq 0$. Thus,
$z(\lambda_2^{**})+\lambda_2^{**} z'(\lambda_2^{**}) \geq 0$. By the mean value
theorem, there exists $\hat{\lambda}_2\in(\lambda_2^{**}, 1)$ such that
$z(\lambda_2^{**})=z(\lambda_2^{**})-z(1)=z'(\hat{\lambda}_2)(\lambda_2^{**}-1)$.
Then we have
\begin{align}
\label{AppendixCLong_1} 0 &\leq z(\lambda_2^{**})+\lambda_2^{**}
z'(\lambda_2^{**})\\ &= z'(\hat{\lambda}_2)(\lambda_2^{**}-1)
+ \lambda_2^{**} z'(\lambda_2^{**})\\
&\leq z'(\lambda_2^{**})(\lambda_2^{**} - 1)
+\lambda_2^{**}z'(\lambda_2^{**})\\
& = z'(\lambda_2^{**})(2\lambda_2^{**} - 1),
\end{align}
which implies $\lambda_2^{**} \leq 1/2$.\hfill$\blacksquare$

\section{Proof of Proposition \ref{ExistenceOfEquilibriumPointDuopoly}}

We consider two cases depending on the relative values of $p_1/q_1$
and $p_2/g(0)$.

\textit{Case 1 ($p_1/q_1\leq p_2/g(0)$):} Let $\lambda_1^*=1-F(p_1/q_1)$
and $\lambda_2^*=0$. Since $p_1/q_1 \leq p_2/g(\lambda_2^*)$, $h_{d,1}(\lambda_1^*,
\lambda_2^*)$ and $h_{d,2}(\lambda_1^*,\lambda_2^*)$ are determined by
(\ref{EvolutionDuopolyS1_2}) and (\ref{EvolutionDuopolyS2_2}), respectively.
Thus, $h_{d,1}(\lambda_1^*,\lambda_2^*) = \lambda_1^*$ and
$h_{d,2}(\lambda_1^*,\lambda_2^*) = \lambda_2^*$, and by Definition 3,
$(\lambda_1^*,\lambda_2^*)$ is an equilibrium point.
By the non-increasing property of $g(\cdot)$, we have $p_1/q_1\leq p_2/g(0) \leq p_2/g(\lambda_2)$
for all $\lambda_2\in[0,1]$. Thus, the user subscription dynamics in Case 1
is described by (\ref{EvolutionDuopolyS1_2}) and (\ref{EvolutionDuopolyS2_2}).
This establishes the uniqueness of the equilibrium point $(\lambda_1^*,\lambda_2^*)$,
because $h_{d,i}(\cdot)$ cannot take a value different from $\lambda_i^*$, for $i=1,2$.

\textit{Case 2 ($p_1/q_1 > p_2/g(0)$):} Since $h_{d,2}(\cdot)$ is independent of $\lambda_1$, we can
express $h_{d,2}(\cdot)$ as a function of $\lambda_2$ only:
\begin{align}
h_{d,2}(\lambda_2) = \left\{
\begin{array}{ll}
F\left(\frac{p_1-p_2}{q_1-g(\lambda_2) }\right) -
F\left(\frac{p_2}{g(\lambda_2)}\right) \quad &\text{if } \frac{p_1}{q_1} > \frac{p_2}{g(\lambda_2)},\\
0 &\;\;\;\;\;\;\text{otherwise}.
\end{array} \right.
\end{align}
Note that $h_{d,2}(\cdot)$ is continuous and non-increasing on
$[0,1]$. Let $\tilde{h}_{d,2}(\lambda_2) = h_{d,2}(\lambda_2) -
\lambda_2$ for all $\lambda_2 \in [0,1]$. Then
$\tilde{h}_{d,2}(\cdot)$ is continuous and strictly decreasing on
$[0,1]$. Since $p_1/q_1 > p_2/g(0)$, we have
\begin{align}
\tilde{h}_{d,2}(0) = h_{d,2}(0) - 0 = F\left(\frac{p_1-p_2}{q_1-g(0) }\right) - F\left(\frac{p_2}{g(0)}\right) > 0.
\end{align}
Also,
\begin{align}
\nonumber
\tilde{h}_{d,2}(1) &= h_{d,2}(1) - 1 \\
&= \left\{
\begin{array}{ll}
F\left(\frac{p_1-p_2}{q_1-g(1) }\right) -
F\left(\frac{p_2}{g(1)}\right) - 1 \quad &\text{if } \frac{p_1}{q_1} > \frac{p_2}{g(1)},\\
- 1 &\;\;\;\;\text{otherwise}.
\end{array} \right.
\end{align}
Hence, $\tilde{h}_{d,2}(1) \leq 0$, and there exists unique $\lambda_2^* \in (0,1]$
such that $h_{d,2}(\lambda_2^*) = \lambda_2^*$.
Suppose that $p_1/q_1\leq p_2/g(\lambda_2^*)$. Then $\lambda_2^* = h_{d,2}(\lambda_2^*) = 0$,
which is a contradiction. Hence, $\lambda_2^*$ must satisfy $p_1/q_1 > p_2/g(\lambda_2^*)$.
Let $\lambda_1^* = F\left((p_1-p_2)/(q_1-g(\lambda_2^*))\right)$. Then it is easy
to verify that $(\lambda_1^*,\lambda_2^*)$ is an equilibrium point. This equilibrium
point is unique because $\lambda_2^*$ is the unique fixed point of $h_{d,2}(\cdot)$.\hfill$\blacksquare$

\section{Proof of Theorem \ref{ConvergenceOfEvolutionDuopoly}}

For notational convenience, we use $h_{d,1}(\lambda_2)$ and
$h_{d,2}(\lambda_2)$ instead of $h_{d,1}(\lambda_1,\lambda_2)$ and
$h_{d,2}(\lambda_1,\lambda_2)$, respectively, since they are
independent of $\lambda_1$. Note that $h_{d,1}(\cdot)$ is a
non-decreasing function of $\lambda_2$ while $h_{d,2}(\cdot)$ is a
non-increasing function of $\lambda_2$. Define a mapping
$\mathbf{h}_{d}:\Lambda\to\Lambda$ by
\begin{equation}
\label{MappingDefinitionAppendixE}
\mathbf{h}_{d}(\lambda_1,\lambda_2)=\left(h_{d,1}(\lambda_2),
h_{d,2}(\lambda_2) \right).
\end{equation}
To prove Theorem \ref{ConvergenceOfEvolutionDuopoly}, we shall
show that the mapping $\mathbf{h}_{d}(\cdot)$ is a contraction
on $\Lambda$ with respect to the maximum norm \cite{BertsekasTsitsiklis}
if the condition \eqref{ConvergenceConditionDuopoly} is satisfied.

Let $\lambda_a=(\lambda_{1,a},\lambda_{2,a})$ and
$\lambda_b=(\lambda_{1,b},\lambda_{2,b})$ be two different vectors
arbitrarily chosen from the set
$\Lambda$, and suppose without loss of generality that $\lambda_{2,a}\geq \lambda_{2,b}$.
We will show that
\begin{eqnarray}
\label{ContractionDuopoly}
\|\mathbf{h}_{d}(\lambda_{1,a},\lambda_{2,a})-\mathbf{h}_{d}(\lambda_{1,b},\lambda_{2,b})\|_{\infty} \leq
\kappa_d\left\|\lambda_{a}-\lambda_{b}\right\|_{\infty},
\end{eqnarray}
where $\kappa_d=K\cdot\max_{\lambda_2\in[0,1]}\left\{[-g'(\lambda_2)/g(\lambda_2)]
\cdot [q_1/(q_1-g(\lambda_2))]\right\}$. Then the condition \eqref{ConvergenceConditionDuopoly}
implies that $\kappa_d \in [0,1)$, establishing that $\mathbf{h}_{d}(\cdot)$ is a contraction.
Since $p_2/g(\lambda_{2,b}) \leq p_2/g(\lambda_{2,a})$, we can
consider three cases.

\textit{Case 1 ($p_2/g(\lambda_{2,a}),  p_2/g(\lambda_{2,b}) < p_1/q_1$):}
In this case, both $\mathbf{h}_{d}(\lambda_{1,a},\lambda_{2,a})$ and $\mathbf{h}_{d}(\lambda_{1,b},\lambda_{2,b})$ are determined by
(\ref{EvolutionDuopolyS1}) and (\ref{EvolutionDuopolyS2}). Hence,
\begin{eqnarray}
 \label{long1}
& &\|\mathbf{h}_{d}(\lambda_{1,a},\lambda_{2,a})-\mathbf{h}_{d}(\lambda_{1,b},\lambda_{2,b})\|_{\infty}\\
 \label{long2}
&=&\max_{i=1,2} \{ \left|h_{d,i}(\lambda_{2,a})-h_{d,i}(\lambda_{2,b})\right| \}\\
\nonumber &=&\max\left\{F\left(\frac{p_1-p_2}{q_1-g(\lambda_{2,b})
}\right)-F\left(\frac{p_1-p_2}{q_1-g(\lambda_{2,a})
}\right)\right.,\\
 \nonumber&
&\;\;\;\;\;\;\;\;\;F\left(\frac{p_1-p_2}{q_1-g(\lambda_{2,b})
}\right)-F\left(\frac{p_1-p_2}{q_1-g(\lambda_{2,a}) }\right)\\
\label{long3}& & \left.\;\;\;\;\;\;\;\;\; +
F\left(\frac{p_2}{g(\lambda_{2,a}) }\right)
- F\left(\frac{p_2}{g(\lambda_{2,b}) }\right)\right\}\\
\label{long4} &=& h_{d,2}(\lambda_{2,b})-h_{d,2}(\lambda_{2,a})\,.
\end{eqnarray}

Note that \eqref{ContractionDuopoly} is trivially satisfied if $\lambda_{2,a}=\lambda_{2,b}$.
Thus, we assume that $\lambda_{2,a} > \lambda_{2,b}$.
We first consider the case where $0 < p_2/g(\lambda_{2,b}) \leq p_2/g(\lambda_{2,a}) < \beta$
and $0 < (p_1-p_2)/(q_1-g(\lambda_{2,a})) \leq (p_1-p_2)/(q_1-g(\lambda_{2,b})) < \beta$
so that $h_{d,2}(\cdot)$ is differentiable on $(\lambda_{2,b},\lambda_{2,a})$.
Then by the mean value theorem, there exists $\lambda_{2,c}\in(\lambda_{2,b}, \lambda_{2,a})$
such that $h_{d,2}(\lambda_{2,b})-h_{d,2}(\lambda_{2,a}) = h'_{d,2}(\lambda_{2,c}) (\lambda_{2,b} - \lambda_{2,a})$.
Therefore,
\begin{align}
&\left\|\mathbf{h}_{d}(\lambda_{1,a},\lambda_{2,a})-\mathbf{h}_{d}(\lambda_{1,b},\lambda_{2,b})\right\|_{\infty}\\
=&\Bigg\{-f\left(\frac{p_1-p_2}{q_1-g(\lambda_{2,c})}\right)\frac{\left(p_1-p_2\right)g'(\lambda_{2,c})}{\left[q_1-g(\lambda_{2,c})\right]^2}\\
&-f\left(\frac{p_2}{g(\lambda_{2,c})}\right)\frac{p_2g'(\lambda_{2,c})}{(g(\lambda_{2,c}))^2}\Bigg\} \left|\lambda_{2,a}-\lambda_{2,b}\right|\\
\label{long7} \leq& \ K \left[-\frac{g'(\lambda_{2,c})}{q_1-g(\lambda_{2,c})}
-\frac{g'(\lambda_{2,c})}{g(\lambda_{2,c})}\right] \left|\lambda_{2,a}-\lambda_{2,b}\right|\\
\label{long8}
=& \ K \left[-\frac{g'(\lambda_{2,c})}{g(\lambda_{2,c})}
\cdot\frac{q_1}{q_1-g(\lambda_{2,c})}\right] \left|\lambda_{2,a}-\lambda_{2,b}\right|\\
\label{long9}
\leq& \ \kappa_d \left\|\lambda_{a}-\lambda_{b}\right\|_{\infty}.
\end{align}
The cases where $0 < p_2/g(\lambda_{2,b}) \leq p_2/g(\lambda_{2,a}) < \beta$
or $0 < (p_1-p_2)/(q_1-g(\lambda_{2,a})) \leq (p_1-p_2)/(q_1-g(\lambda_{2,b})) < \beta$
does not hold can be covered as in the proof of Theorem \ref{ConvergenceOfEvolution}.

\textit{Case 2 ($p_2/g(\lambda_{2,b}) < p_1/q_1 \leq p_2/g(\lambda_{2,a})$):}
In this case, $\mathbf{h}_{d}(\lambda_{1,a},\lambda_{2,a})$ is determined by
(\ref{EvolutionDuopolyS1_2}) and (\ref{EvolutionDuopolyS2_2}), while
$\mathbf{h}_{d}(\lambda_{1,b},\lambda_{2,b})$ is determined by
(\ref{EvolutionDuopolyS1}) and (\ref{EvolutionDuopolyS2}).
Hence,
\begin{align}
 \label{long1_2}
&\|\mathbf{h}_{d}(\lambda_{1,a},\lambda_{2,a})-\mathbf{h}_{d}(\lambda_{1,b},\lambda_{2,b})\|_{\infty}\\
\label{long3_2}
=&\max\left\{F\left(\frac{p_1-p_2}{q_1-g(\lambda_{2,b})}\right)-F\left(\frac{p_1}{q_1}\right),\right.\\
&\left.
\;\;\;\;\;\;\;\;\;F\left(\frac{p_1-p_2}{q_1-g(\lambda_{2,b})}\right)-F\left(\frac{p_2}{g(\lambda_{2,b})}\right)
\right\}\\
\label{long4_2} =&F\left(\frac{p_1-p_2}{q_1-g(\lambda_{2,b})}\right)
-
F\left(\frac{p_2}{g(\lambda_{2,b})}\right)\\
\label{long5_2} =&h_{d,2}(\lambda_{2,b})-h_{d,2}(\lambda_{2,a})
\end{align}
Let $\bar{\lambda}_2 = \min \{ \lambda_2 \in [0,1] | \ g(\lambda_2)
p_1/q_1 \leq p_2 \}$. Note that $\lambda_{2,b} < \bar{\lambda}_2
\leq \lambda_{2,a}$. Again, we focus on the case where $0 <
p_2/g(\lambda_{2,b}) \leq p_2/g(\bar{\lambda}_2) < \beta$ and $0 <
(p_1-p_2)/(q_1-g(\bar{\lambda}_2)) \leq
(p_1-p_2)/(q_1-g(\lambda_{2,b})) < \beta$,  while omitting the other
cases. Applying the mean value theorem to $h_{d,2}(\cdot)$ on the
interval $[\lambda_{2,b}, \bar{\lambda}_2]$ yields
\begin{equation}
\label{InequalityAppendixF2}
|h_{d,2}(\bar{\lambda}_2)-h_{d,2}(\lambda_{2,b})| \leq \kappa_d |\bar{\lambda}_2-\lambda_{2,b}|.
\end{equation}
Since $h_{d,2}(\bar{\lambda}_2) = h_{d,2}(\lambda_{2,a}) = 0$ and $\kappa_d \geq 0$,
we obtain
\begin{align}
&\|\mathbf{h}_{d}(\lambda_{1,a},\lambda_{2,a})-\mathbf{h}_{d}(\lambda_{1,b},\lambda_{2,b})\|_{\infty}\\
=& |h_{d,2}(\lambda_{2,a})-h_{d,2}(\lambda_{2,b})|\\
=& |h_{d,2}(\bar{\lambda}_2)-h_{d,2}(\lambda_{2,b})| \\
\leq& \kappa_d |\bar{\lambda}_2-\lambda_{2,b}| \leq \kappa_d
|\lambda_{2,a}-\lambda_{2,b}|\\
\leq & \kappa_d \left\|\lambda_{a}-\lambda_{b}\right\|_{\infty}.
\end{align}

\textit{Case 3 ($p_2/g(\lambda_{2,a}),  p_2/g(\lambda_{2,b}) \geq p_1/q_1$):}
In this case, both $\mathbf{h}_{d}(\lambda_{1,a},\lambda_{2,a})$ and $\mathbf{h}_{d}(\lambda_{1,b},\lambda_{2,b})$ are determined by
(\ref{EvolutionDuopolyS1_2}) and (\ref{EvolutionDuopolyS2_2}). Hence,
$\mathbf{h}_{d}(\lambda_{1,a},\lambda_{2,a})=\mathbf{h}_{d}(\lambda_{1,b},\lambda_{2,b})=(1-F(p_1/q_1),
0)$, and \eqref{ContractionDuopoly} is trivially satisfied.\hfill$\blacksquare$

\section{Proof of Theorem \ref{ExistenceOfNashEquilibriumDuopoly}}

Let $\tilde{\mathcal{G}}'_C=\left\{\mathcal{S}_i,
R_i(\lambda_1,\lambda_2), \lambda_i\in [0,1/2] \;|\;i=1,2\right\}$.
Since $f(\cdot)$ is non-increasing, the set of NE of
$\tilde{\mathcal{G}}_C$ is equal to that of $\tilde{\mathcal{G}}'_C$
by Lemma \ref{SumMarketShareLessThanOne}, and thus it suffices to
prove the existence of NE of $\tilde{\mathcal{G}}'_C$. Since
$[0,1/2]^2 \subset \Lambda$, the expression of
$R_i(\lambda_1,\lambda_2)$ in $\tilde{\mathcal{G}}'_C$ is given by
$R_i(\lambda_1,\lambda_2) = \lambda_i p_i(\lambda_1,\lambda_2)$, for
$i=1,2$. Since $f(\cdot)$ is continuously differentiable on
$[0,\beta]$, $R_i(\cdot)$ is twice continuously differentiable on
$[0,1/2]^2$. The second-order mixed partial derivatives of
$R_1(\cdot)$ and $R_2(\cdot)$ are given by the negative of the
left-hand sides of (\ref{SupermodularCondition_1}) and
(\ref{SupermodularCondition_2}), respectively. Therefore,
(\ref{SupermodularCondition_1}) and (\ref{SupermodularCondition_2})
imply that $\frac{\partial^2
R_i(\lambda_1,\lambda_2)}{\partial\lambda_1\partial\lambda_2} \leq
0$,
for all $(\lambda_1,\lambda_2) \in [0,1/2]^2$, for all $i=1,2$.
Consider a transformed game of $\tilde{\mathcal{G}}'_C$,
\begin{equation}
\begin{split} \tilde{\mathcal{G}}''_C=&\left\{\mathcal{S}_1,
\mathcal{S}_2,\tilde{R}_1(\mu_1,\lambda_2),
\tilde{R}_2(\mu_1,\lambda_2),\right.\\
& \;\;\;\;\; \mu_1 \in [-1/2,0], \lambda_2\in [0,1/2] \Big{\}},
\end{split}
\end{equation}
where $\tilde{R}_i(\mu_1,\lambda_2) = R_i(-\mu_1,\lambda_2)$
for all $(\mu_1,\lambda_2) \in [-1/2,0]\times[0,1/2]$, for $i=1,2$.
Since $\tilde{R}_1(\cdot)$ and $\tilde{R}_2(\cdot)$ are continuous
in both arguments and have increasing differences in $(\mu_1,\lambda_2)$
and in $(\lambda_2,\mu_1)$, respectively,
$\tilde{\mathcal{G}}''_C$ is a supermodular game \cite{Topkis}. Therefore, $\tilde{\mathcal{G}}''_C$
has at least one pure NE, and $\tilde{\mathcal{G}}'_C$ also has a corresponding NE.\hfill $\blacksquare$


\end{document}